\documentclass[10pt,svgnames]{iopart}

\usepackage{graphicx}
\usepackage{dcolumn}
\usepackage{bm}

\usepackage[utf8]{inputenc}
\usepackage[greek,english]{babel}
\usepackage[autostyle]{csquotes}
\usepackage{textgreek}

\usepackage{siunitx}
\DeclareSIUnit{\arbitraryunit}{arb.~u.}
\DeclareSIUnit{\photons}{photons}
\DeclareSIUnit{\counts}{counts}
\DeclareSIUnit{\pixel}{pixel}
\DeclareSIUnit{\torr}{torr}
\DeclareSIUnit{\sccm}{sccm}
\usepackage{todonotes}
\expandafter\let\csname equation*\endcsname\relax
\expandafter\let\csname endequation*\endcsname\relax
\usepackage{amssymb}
\usepackage[version=4]{mhchem}
\usepackage[style=phys,biblabel=brackets,backend=biber,citestyle=numeric-comp,sorting=none,doi=false,isbn=false]{biblatex}
\usepackage{booktabs}					

\definecolor{blue}{rgb}{0,0.447,0.741}
\definecolor{orange}{rgb}{0.85,0.325,0.098}
\definecolor{yellow}{rgb}{0.929,0.694,0.125}
\definecolor{violet}{rgb}{0.494,0.184,0.556}

\begin{document}
	
\title[Energy transport in an atmospheric pressure plasma jet]{Multi-diagnostic approach to energy transport in an atmospheric pressure helium-oxygen plasma jet}

\author{Tristan Winzer$^{1}$, Natascha Blosczyk$^{1}$, Jan Benedikt$^1$, Judith Golda$^{1,2}$}
\address{$^{1}$Institute of Experimental and Applied Physics, Kiel University, Germany.}
\address{$^{2}$Plasma Interface Physics, Ruhr-University Bochum, Germany.}

\ead{winzer@physik.uni-kiel.de}
\vspace{10pt}
\begin{indented}
		\item[]\date{\today}
\end{indented}
	
\begin{abstract}
The energy balance of a plasma holds fundamental information not only about basic plasma physics, but it is also important for tailoring plasmas to specific applications. Especially RF-driven atmospheric pressure plasma jets (APPJs) operated in helium with oxygen admixture have high application potential in industry and medicine. Many types of plasma jets have been studied up to now, leading to the challenge how to compare results from various sources. We have developed a method for measuring the power deposited in the plasma as the parameter to compare different sources and gas mixtures with each other. Furthermore, we studied energy transport as a function of this input power and molecular gas admixture in a newly developed APPJ based on the COST-reference jet with a capillary as a dielectric in between the electrodes. The gas temperature, atomic oxygen density, ozone density and absolute emission intensity in the visible wavelength range have been determined. Combining the results gave an energy balance with most of the energy deposited into gas heating. Production of final chemical products made up a small amount of the deposited power while radiation was negligible for all combinations of external parameters studied.
\end{abstract}

\vspace{2pc}
\noindent{\it Keywords}: atmospheric pressure plasma, \textmu-APPJ, dissipated power, optical emission spectroscopy, oxygen actinometry, molecular beam mass spectrometry
%
%
%
\ioptwocol
%
	
	
\section{Introduction}

Atmospheric pressure plasma jets (APPJs) have gained increased interest in recent years, due to their application potential in medicine and industrial processes \cite{Kogelschatz.2004,Lazzaroni.2012,Ehlbeck.2011,Graves.2012,Park.2012}. These plasmas exhibit strong non-equilibrium characteristics with high electron energies at low gas temperature, especially when driven with RF-frequencies. The energy is in this case selectively coupled to the electrons and is then dissipated via different physical processes to other species present in the plasma, e.g. via drift, diffusion, elastic and inelastic collisions. To channel the energy inside the discharge into the desired processes, fundamental understanding of the influence of external parameters on the energy balance is necessary.

An energy balance answers the question of how the energy introduced into the plasma system is distributed among the species and processes. In history, the Lawson criterion is probably one of the most prominent examples for successful consideration of an energy balance for a plasma in thermal equilibrium \cite{Lawson.1957}. As early as 1957, \citeauthor{Lawson.1957} theoretically compared the energy produced in a fusion reaction with the energy lost to the environment. In doing so, he defined a minimum required value for the product of the plasma density and the confinement time that results in a net energy output. Another example for a thorough analysis of an energy balance is the model of equi-operational plasmas by \citeauthor{Jonkers.2002} from 2003, where the particle balance, the energy balance, and the relative population of the ground state were used to derive three characteristic plasma parameters: characteristic size, power density and pressure. When these parameters are identical, two plasmas using different feed gas mixtures are comparable \cite{Jonkers.2002}.

Here, we experimentally studied the energy balance of an atmospheric pressure plasma using an RF-driven capacitively coupled APPJ with a capillary as dielectric between the electrodes. The first step was to measure how much energy is delivered to the plasma. The following analysis focuses on the loss mechanisms of the electron energy. We took into account elastic and inelastic collisions. Losses by elastic collisions of electrons with ions and neutrals are transferred into translational energy of these atoms and molecules, i.e. heat. This energy fraction leaving the plasma with the gas flow is represented by the neutral gas temperature. Inelastic collisions lead to excitation, ionization and dissociation. Part of this energy is initiating chemical reactions in the plasma, for example the generation of atomic oxygen. The excitation is represented by quantitatively characterizing the radiation emitted during de-excitation.

\section{Experimental setup}
\label{sec:exp}
	
\subsection{Atmospheric pressure plasma jet}
\label{sec:capillary_jet}

For studying the influence of different external parameters on the energy transport in the plasma, a setup which ensures stable plasma operation over a broad range of parameters and on long timescales is needed. As reproducibility is also very important for comparing the results of different measurement series, we employ a derivative of the COST Reference Microplasma Jet. The COST-Jet has shown to deliver reproducible and stable operating conditions. However, it is limited in terms of the maximum input power and molecular gas admixture. At input powers between \SI{1.5}{\watt} and \SI{2}{\watt}, the COST-Jet switches to a constricted discharge mode at the tip of the electrodes leading to possible thermal damage \cite{Golda.2016, Golda.2019}. At molecular gas admixtures above \SI{1.6}{\percent}, the plasma in the COST-Jet extinguishes \cite{Ellerweg.2010}.

\begin{figure}[!ht]
	\centering
	\includegraphics[width=0.99\linewidth]{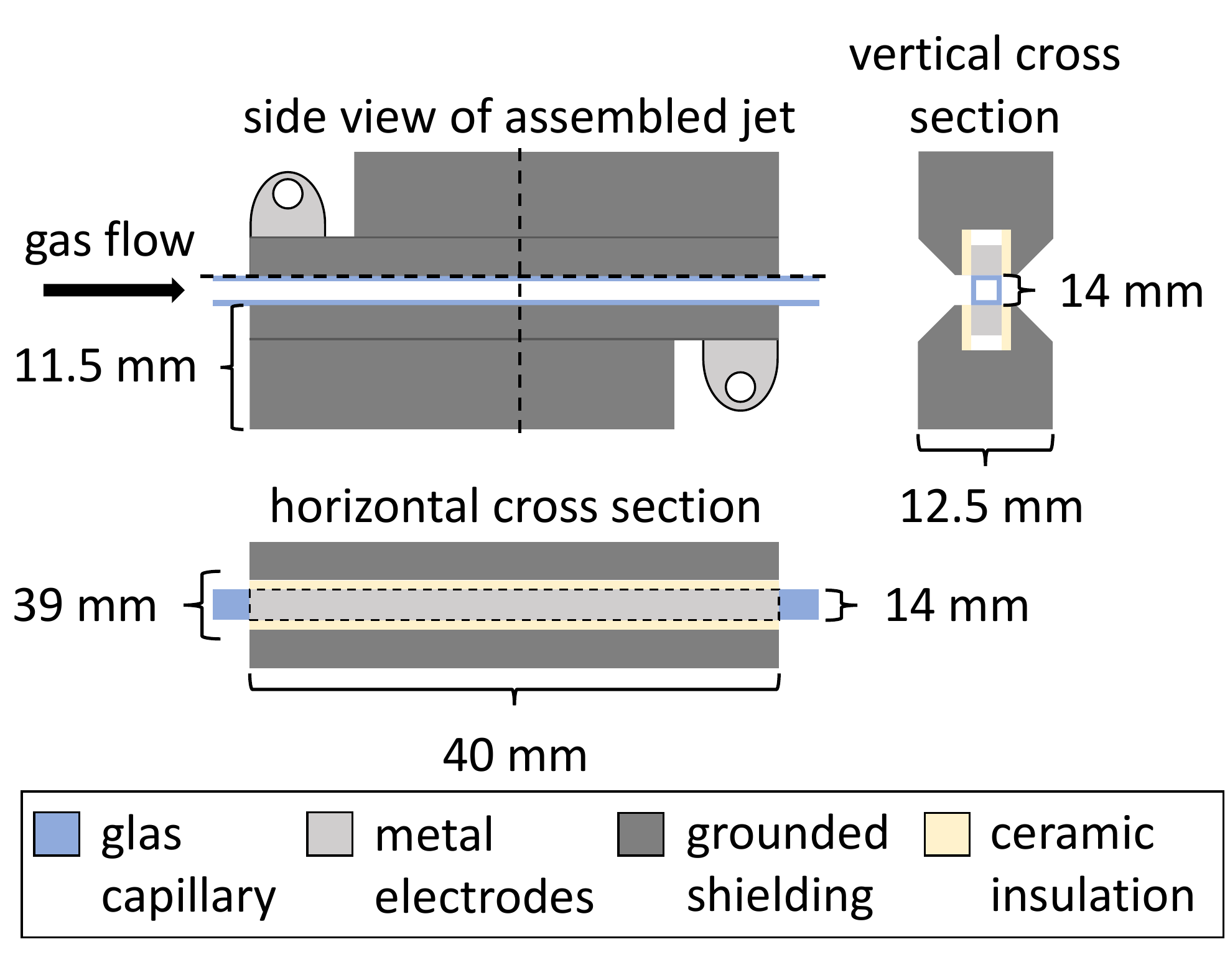}
	\caption{Schematic drawing of the capillary jet device. The side view of the jet is shown in the upper left corner. The two other views are the cross sections at the positions of the dashed lines.}
	\label{fig:cap_jet}
\end{figure}

The so called capillary plasma jet features a similar discharge geometry and excitation scheme as the COST-Jet \cite{Golda.2016}, see Figure~\ref{fig:cap_jet}. The plasma is generated inside a glas capillary with a quadratic inner cross section of \SI{1}{\milli\meter\squared} and a wall thickness of \SI{0.2}{\milli\meter}. This leads to a total gap size of \SI{1.4}{\milli\meter} between the two \SI{40}{\milli\meter} long plane-parallel electrodes. The resulting plasma volume is \SI{40}{\cubic\milli\meter}. One of the electrodes is grounded and the other one is connected to the RF-generator via a matching network. Both stainless steel electrodes are held in place by insulating ceramic plates on two sides which are fixed by grounded stainless steel metal shielding. Electrical probes for voltage and current are implemented by small circuit boards screwed to the grounded metal shielding. The voltage probe consists of a pick-up antenna near the powered electrode. Current is measured using a \SI{4.7}{\ohm} current sensing resistor between the grounded electrode and ground. Details about the probes and the electric circuit can be found in the next section.

Gas mixtures used in this work consisted of \SI{1000}{\sccm} helium (\SI{99.999}{\percent}) and \SIrange{0}{20}{\sccm} oxygen (\SI{99.9995}{\percent}) with small admixtures of argon and neon (\SIrange{0.4}{0.5}{\sccm}), when needed for diagnostic purposes. To avoid the intrusion of impurities into the gas flow, we used mainly stainless steel tubing. Only a short section right before the jet consisted of Viton tubing for enhanced flexibility in positioning the plasma device.

\subsection{Power supply and power measurement}
\label{sec:power}

\begin{figure}[!ht]
	\centering
	\includegraphics[width=0.99\linewidth]{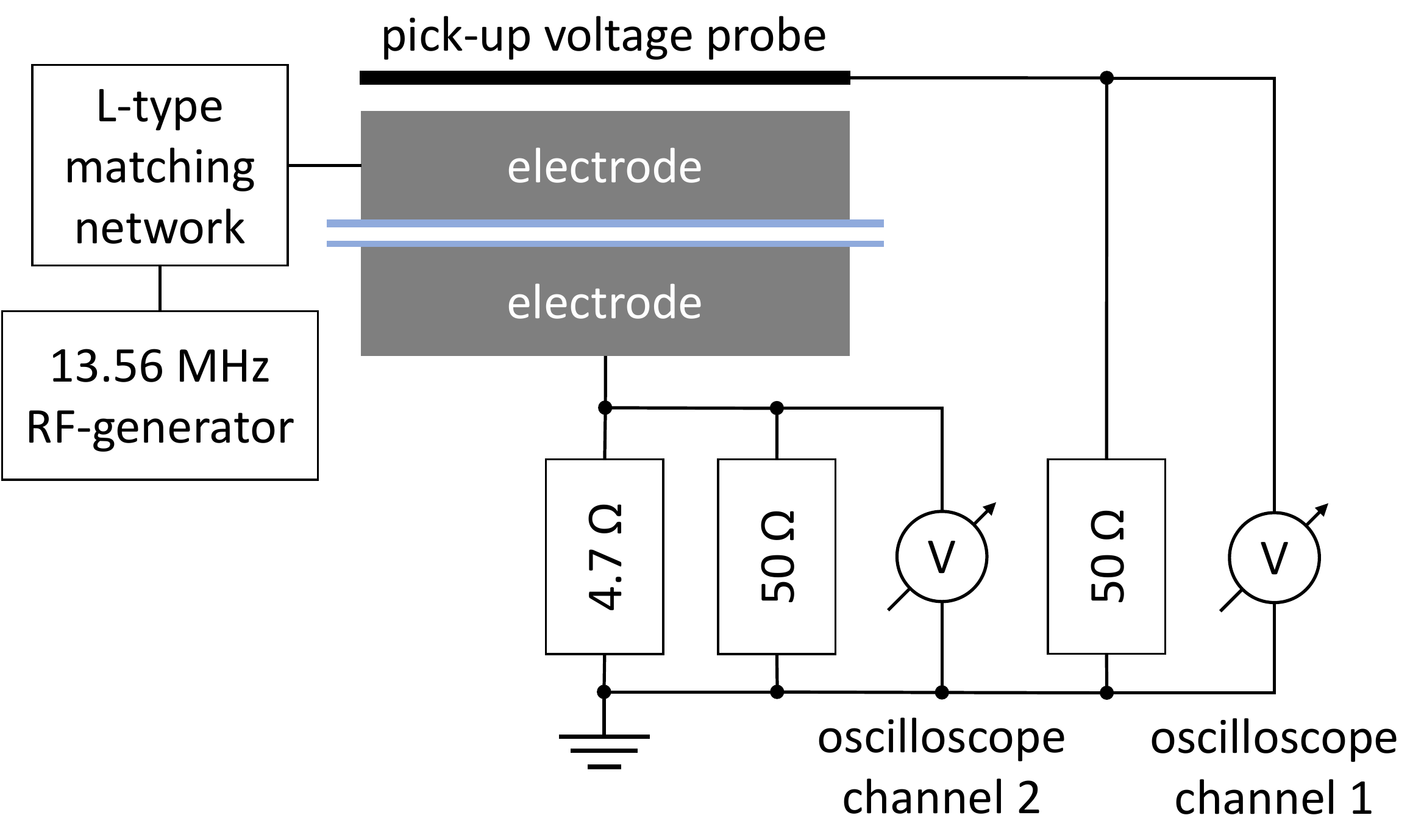}
	\caption{Schematic diagram of the electrical setup used for power measurements.}
	\label{fig:elect_setup}
\end{figure}

Because of losses e.g. in the matchbox or cables, the generator power is always larger than the deposited power. A setup for simultaneously measuring the voltage and current signals at the plasma jet with high temporal resolution is needed to derive the power delivered to the plasma. The setup is adapted from the COST-Jet, where it has delivered stable and reproducible results for the discharge power \cite{Golda.2019}. A sketch of the setup is shown in figure~\ref{fig:elect_setup}.

The voltage is supplied to one electrode by a \SI{13.56}{\mega\hertz} generator (dressler Cesar) connected to a L-type matching network (Barthel Matching Cube i-300). This voltage is measured using a pick-up antenna and a \SI{2.5}{GS/s} oscilloscope (Yokogawa DL9140l). Calibration of the pick-up voltage was performed using a commercial voltage probe (Tektronix P5100A). A second voltage is measured over a \SI{4.7}{\ohm} resistor between ground and the grounded electrode. The voltage drop over the resistor is directly proportional to the current flowing through the plasma via Ohm's law. Two \SI{50}{\ohm} termination resistors are connected in parallel to the input channels of the oscilloscope for minimizing standing wave effects.

The average power supplied to the plasma $P_\text{avg}$ can be calculated from the measured voltage and current signals. For sinusoidal voltages, this can be simplified to

\begin{equation}
    P_\text{avg} = U_\text{rms} I_\text{rms} \cos(\Delta \varphi) \text{ ,}
    \label{eq:average_power}
\end{equation}

where $U_\text{rms}$ and $I_\text{rms}$ are the root-mean-square values of the voltage and current respectively and $\Delta \varphi$ is the phase shift between them. Therefore, an established technique to calculate the power is to fit sine waves to the measured waveforms and extract the necessary values. However, at higher voltages, the processes in the plasma become increasingly nonlinear, leading to distortion of the measured signals \cite{Park.2001}. Consequently, for measurements in higher power regimes, it is necessary to account for these distortions. Therefore, we developed a more elaborate technique based on the cross-correlation method to enable power calculation by point-wise multiplication of voltage and current waveforms.

First, the phase shift between the probes measuring voltage and current has to be determined. For a perfect capacitor, this phase difference is \SI{90}{\degree}. Most capacitively coupled setups without plasma do not resemble perfect capacitors. This has to be accounted for by measuring a reference phase shift $\Delta \varphi_\text{ref}$ without plasma. A powerful method for calculating the time shift between two arbitrary signals is the cross-correlation method. This time shift $\Delta t_\text{ref}$ is directly related to the phase shift $\Delta \varphi_\text{ref}$.

Calculating the cross correlation coefficient $\Psi_\text{UI}$ for two infinite, continuous, sinusoidal signals analytically leads to a cosine-shaped function

\begin{equation}
    \Psi_\text{UI}(\tau) = U_\text{rms} I_\text{rms} \cos(\omega(\Delta t - \tau)) \text{ .}
    \label{eq:correlation_coefficient}
\end{equation}

The shift $\Delta t_\text{ref}$ between the two signals can be calculated from the lag $\tau_\text{max}$ of the maximum correlation coefficient closest to zero lag, because the cosine has its highest value at $\tau = \Delta t$. In reality, the measured waveforms are finite and discretely sampled by the oscilloscope. The so-called unbiased correction of the cross-correlation compensates for the finite nature. To overcome the limitation of the accuracy of $\Delta t_\text{ref}$ due to the time resolution of the oscilloscope, a second order polynomial fit to the maximum correlation coefficient near zero lag is used. The lag $\tau_\text{max}$ is calculated using the vertex formula with the coefficients of the fit. After correcting the time shift between voltage and current without plasma by $\Delta t_\text{ref}$, the voltage and current waveforms with plasma can be multiplied point by point to obtain the period-average power supplied to the plasma

\begin{equation}
    P_\text{avg} = \frac{1}{N} \sum_{n = 1}^{N} U_n I_n \text{ ,}
\end{equation}

where $N$ is the number of sampled data points. As the reference time shift from the polynomial fit is more accurate than the time resolution of the sampled signals, the current is linearly interpolated after the correction to allow point-wise multiplication.

\begin{figure}[!ht]
	\centering
	\includegraphics[width=0.99\linewidth]{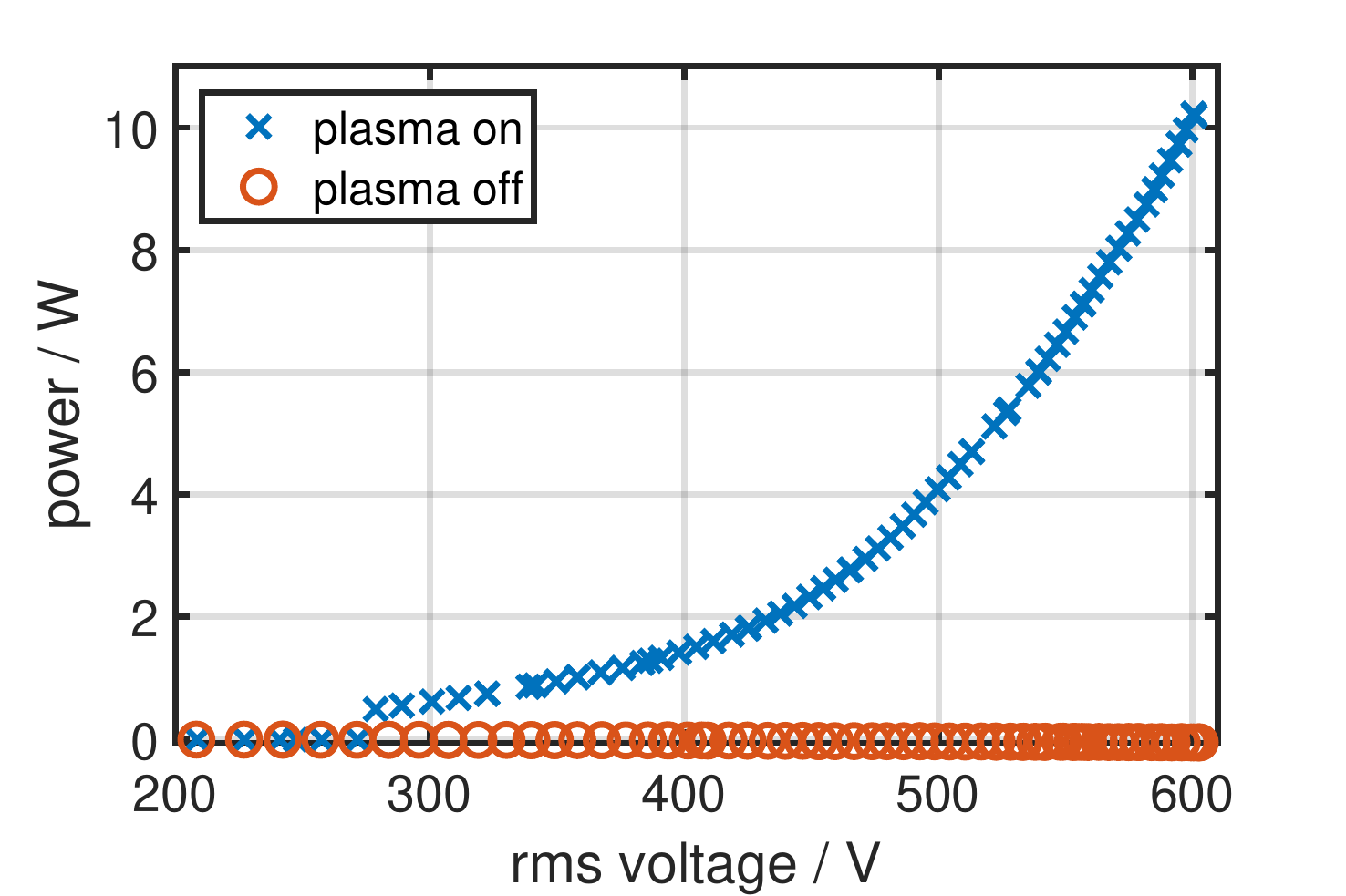}
	\caption{Power coupled to the plasma as a function of the root-mean-square voltage measured with the cross-correlation method. The gas mixture consisted of \SI{1000}{sccm} helium and \SI{5}{sccm} oxygen.}
	\label{fig:power_curve}
\end{figure}

The average plasma input power as a function of the root-mean-square voltage in a range between \SI{200}{\volt_{rms}} and \SI{600}{\volt_{rms}} is shown in blue in figure~\ref{fig:power_curve}. The gas flow consisted of helium with \SI{0.5}{\percent} oxygen. The plasma ignites at approximately \SI{275}{\volt_{rms}}, which is indicated by a jump of the power from \SI{0}{\watt} to \SI{0.3}{\watt}. Afterwards, the power rises linearly with the voltage up to \SI{1.75}{\watt} at \SI{400}{\volt_{rms}}. At higher voltages, the dependence of the power on the voltage changes from linear to quadratic. This change is induced by a change from the so-called $\alpha$-mode to the $\gamma$-mode as already observed and discussed e.g. for the COST-Jet by \citeauthor{Golda.2020b} \cite{Golda.2020b}. A change of the plasma emission from a homogeneous glow in the middle between the electrodes to a bright emission directly at both electrodes is observed in this power region. Ohmic heating of the electrons in the plasma bulk is dominant in the $\alpha$-mode, so it is also referred to as $\Omega$-mode \cite{Hemke.2013}. The $\gamma$-mode is dominated by secondary electrons emitted from the electrodes or created by Penning processes in the sheaths. For this reason, it is also called Penning-mode in the literature  \cite{Bischoff.2018,Gibson.2019}. In contrast to the COST-Jet, which transitions to a constricted discharge at the electrode tips when further increasing the voltage, the dielectric in the capillary jet prevents glow-to-arc transition and the plasma is stable also at high voltages \cite{Shi.2006,Shi.2007}. The maximum measured power in this work is \SI{10}{\watt}. The orange data is the power without the plasma. This curve has been measured with ambient air inside the capillary, which leads to a much higher breakdown voltage. The power is zero over the whole voltage range, confirming that the correction of the time shift is working as expected.

The uncertainty of the measured powers $\Delta P_\text{avg}$ depends partly on the systematic uncertainties of the used resistors and the calibration of the voltage measured by the pick-up antenna to the actual voltage at the electrode. This fraction is calculated using Gaussian error propagation. The other part of $\Delta P_\text{avg}$ depends on the uncertainty of the measured reference phase shift $\Delta (\Delta t_\text{ref})$. It is calculated using the uncertainty of the lag-value $\Delta (\tau_\text{max})$ of the maximum cross correlation coefficient. The uncertainty of the lag is derived from the vertex formula and the polynomial fit coefficients using Gaussian error propagation. The uncertainty values are smaller than the size of the markers in figure~\ref{fig:power_curve} and are therefore not shown here.

\begin{figure}[!ht]
	\centering
	\includegraphics[width=0.99\linewidth]{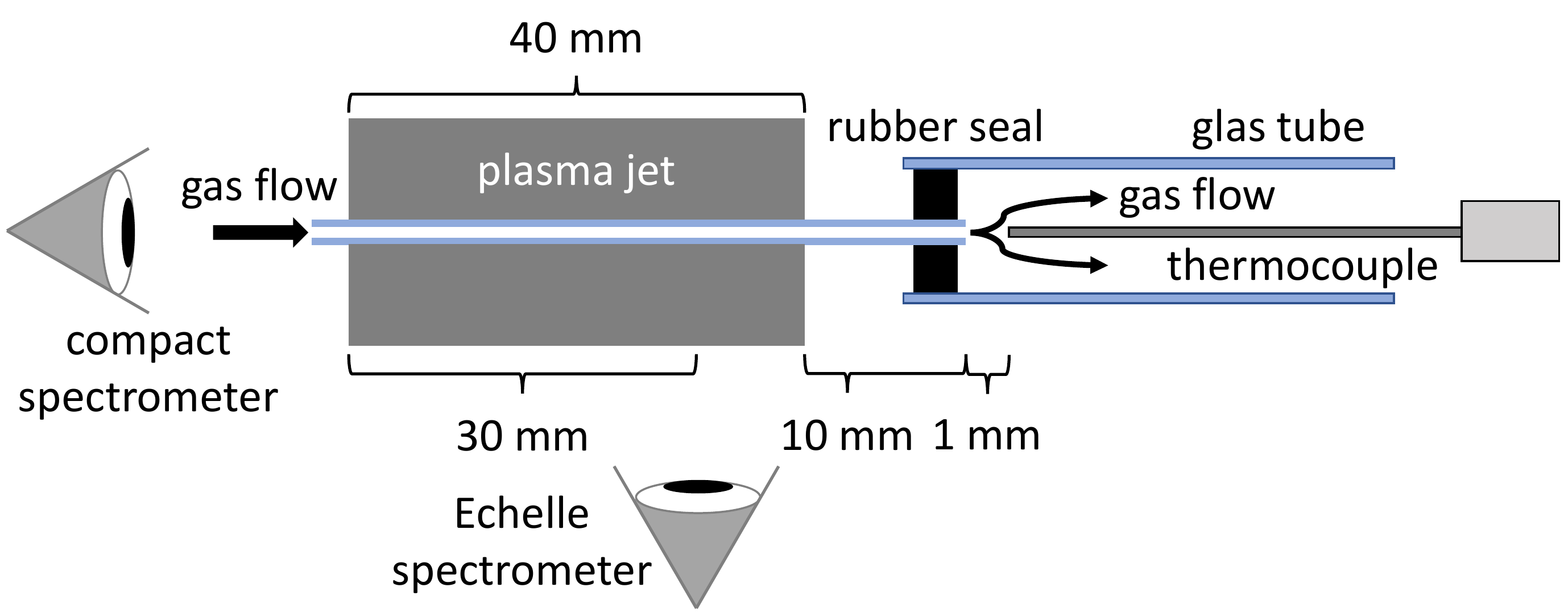}
	\caption{Schematic diagram of the experimental setup used for the optical measurements parallel and perpendicular to the gas flow direction and the measurement of the gas temperature in the effluent.}
	\label{fig:optical_gas_temp_setup}
\end{figure}

\subsection{Optical setup}
\label{sec:opt_setup}

The optical setup used in this work consisted of a combination of two spectrometers measuring the plasma emission parallel and perpendicular to the gas flow direction (figure~\ref{fig:optical_gas_temp_setup}). Parallel to the gas flow, the plasma was imaged into an optical fiber connected to a relatively calibrated spectrometer (Ocean Optics HR4000) by a lens with a focal length of \SI{20}{\milli\meter}.

One-to-one imaging of the plasma perpendicular to the gas flow was performed using two lenses. The first lens had a focal length of \SI{35}{\milli\meter} and the same distance to the plasma. Therefore, the plasma emission was parallelized and afterwards focused into an optical fiber in the focal point of a second lens with a focal length of \SI{50}{\milli\meter}. This optical fiber was connected to an absolutely calibrated high-resolution broadband Echelle spectrometer (LLA Instruments ESA4000 plus).

\begin{figure}
	\centering
	\includegraphics[width=0.99\linewidth]{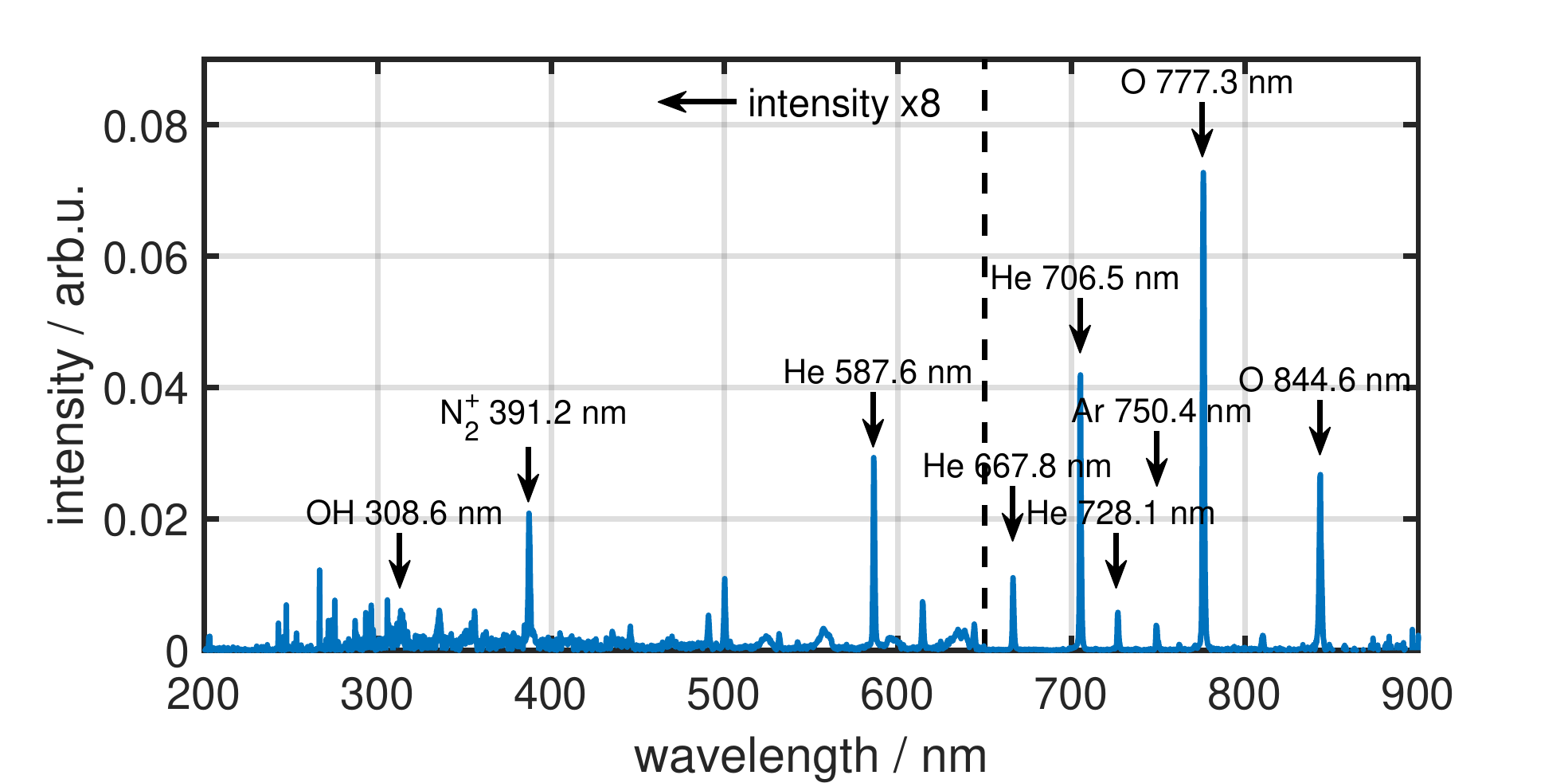}
	\caption{Relatively calibrated spectrum of the plasma for \SI{5}{W} input power and a gas mixture of \SI{1000}{\sccm} helium, \SI{0.4}{\sccm} argon and \SI{5}{\sccm} oxygen taken parallel to the gas flow. The intensities below \SI{650}{\nano\meter} have been scaled up by a factor of \num{8}.}
	\label{fig:spectrum_actinometry}
\end{figure}

An example spectrum of the plasma in $\gamma$-mode with \SI{0.5}{\percent} \ce{O2} and traces of argon taken parallel to the gas flow is shown in figure~\ref{fig:spectrum_actinometry} with the dominant emission features marked by labeled arrows. For better visibility, the intensities below \SI{650}{\nano\meter} have been scaled up by a factor of \num{8}. The emission is dominated by helium and atomic oxygen lines with the only clearly visible argon line at \SI{750.4}{\nano\meter}. The emission below \SI{400}{\nano\meter} can be attributed mostly to nitrogen-impurities from ambient air.

\subsection{Gas temperature}
\label{sec:gas_temp}

As a second non-invasive method, we used thermocouple measurements in the effluent to obtain the gas temperature. The setup is also shown in figure~\ref{fig:optical_gas_temp_setup}. A thermocouple (Type-K) was positioned in front of the capillary and sealed using a glas tube to ensure a laminar gas flow pattern. A rubber seal was positioned around the capillary to avoid ambient intrusion into the glass tube. The signal of the thermocouple was recorded and converted to temperature by an in-house-built measurement electronic originally designed for passive thermal probe measurements \cite{Stahl.2010}. This allowed measurements with \SI{90}{\hertz} repetition rate for investigation of fast changes in the gas temperature. As we could only measure in the effluent about \SI{11}{\milli\meter} away from the plasma, we had to use an extrapolation technique to estimate the temperature in the plasma. With a distance variation, we could extrapolate the measured gas temperature at the end of the electrodes to be approximately \SI{21}{\percent} higher than in the effluent with respect to a room temperature of \SI{20}{\degreeCelsius}.

\subsection{Molecular beam mass spectrometry}
\label{sec:mass_spectrometry}

We used molecular beam mass spectrometry (MBMS) to obtain absolute densities of atomic oxygen and ozone on the central axis of the plasma effluent. Sampling of the species from atmospheric pressure was performed using a setup consisting of three differential pumping stages. Gas enters the first stage through an orifice with a diameter of \SI{50}{\micro\meter}, which has been laser drilled into a molybdenum plate of \SI{250}{\micro\meter} thickness. The first stage is pumped by a scroll pump, while the second and third stage are evacuated using turbo-molecular pumps. A moving chopper with a skimmer was used in the first pumping stage to achieve high beam-to-background ratios and to further reduce the background pressure. This setup is able to reach the necessary UHV-conditions in the third stage housing the mass spectrometer. Details on the setup and the moving skimmer concept can be found in the literature \cite{Benedikt.2009,Benedikt.2013,GroeKreul.2015}. This applies also to the setup and basics of the quadrupole mass spectrometer (HIDEN EPIC1000) used in this study \cite{Benedikt.2012}. Instead of the rotating skimmer concept shown in the literature, we used a linearly moving chopper here.

As we expected gas temperatures of up to a few hundred \si{\degreeCelsius} based on preliminary measurements, a stainless steel cooling plate was mounted onto the orifice plate to ensure constant temperature conditions. This introduced limitations on the jet setup, so sampling of the plasma species through the orifice was performed about \SI{15}{\milli\meter} behind the plasma. To avoid mixing with ambient air and consequently turbulence resulting in enhanced recombination of reactive species, the capillary extended \SI{14}{\milli\meter} of this distance. The resulting gap between the capillary tip and the orifice was \SI{1}{\milli\meter}. This was achieved by moving the electrodes shown in figure~\ref{fig:optical_gas_temp_setup} \SI{4}{\milli\meter} to the left.

Calibration of the measured atomic oxygen signal has been performed by admixing small amounts of neon (\SI{0.5}{sccm}) to the plasma feed gas and simultaneously measuring the neon signal. Knowing the gas temperature, the neon density in front of the orifice can be calculated from the pressure and the flow. This prevents errors due to different plasma conditions during the calibration procedure and the measurement. To avoid creation of atomic oxygen by dissociative ionization of \ce{O2} in the ionizer, we used threshold ionization mass spectrometry (TIMS) with electron energies of \SI{15}{\electronvolt} and \SI{25}{\electronvolt} for atomic oxygen and neon, respectively. Ozone density calibration was realised in the same way by admixing \SI{0.5}{sccm} argon to the feed gas and simultaneously measuring argon and ozone signals at \SI{70}{\electronvolt} electron energy.
	
\section{Results and discussion}
\label{sec:results}

The following discussion of the experimental results is based on the assumption, that the power dissipated in the plasma is coupled to the electrons. This assumption is valid, because the ion with the highest density in \ce{He/O2}-discharges is \ce{O2+}, which has a much lower plasma frequency than the RF-frequency \cite{Lazzaroni.2012}. Energy loss by ions accelerated in the sheaths is also neglected. Due to the high collisionality in the sheath region, the ion energy is directly converted to a temperature increase of the feed gas. The heated electrons distribute the power inside the plasma by elastic and inelastic collisions. Whereas elastic collisions mainly lead to heating of the heavy plasma species and the neutral gas, inelastic collisions lead to excitation, ionization or dissociation and subsequently to radiation and chemical reactions. The power dissipated in these channels under variation of external parameters provides information about the possibility to tailor the energy transport. We chose the total input power and oxygen admixture as the two external parameters, as they have a big influence on electron density and energy distribution. To study the influence of the plasma operating modes on the measured quantities, we varied the input power in a range between \SIrange{0}{6}{\watt} with a constant oxygen admixture of \SI{0.5}{\percent}. For the same reason, the oxygen admixture variation has been performed for constant input powers of \SI{1}{\watt} ($\alpha$-mode) and \SI{5}{\watt} ($\gamma$-mode). Oxygen admixtures ranged from \SIrange{0}{2}{\percent} due to limitations of the setup.
	
\subsection{Heat}
\label{sec:heat}

Although energy transfer of electrons to atoms and molecules by elastic collisions is negligible at low pressures, it has to be taken into account at high pressure due to high collision frequencies. This is why we focused on heat as one significant component of energy transport.

In addition to thermocouple measurements in the effluent with the setup described in section~\ref{sec:gas_temp}, we used the evaluation of molecular rotational spectra to derive the rotational temperature and get a second estimate for the gas temperature in the plasma. The high resolution of the Echelle-spectrometer allowed for identification of individual lines in rotational spectra. For atmospheric pressure plasmas, typically the hydroxyl radical (\ce{OH}) or the nitrogen molecular ion (\ce{N2+}) are used \cite{Bruggeman.2014}. Comparing the rotational temperatures of both species, the temperature determined from the nitrogen ion emission was way higher than the one from \ce{OH}. It is well known, that especially for helium plasmas, the nitrogen levels often interact with those of helium, which leads to an overestimation of the gas temperature \cite{IonascutNedelcescu.2008,Hofmann.2011}. 
Therefore, we decided for the rotational temperature of \ce{OH} as an estimate for the gas temperature and used a bubbler system to admix a controlled amount of water vapor to the feed gas. Details about the bubbler-setup and the control of the water admixture to the plasma feed gas can be found in the literature \cite{Benedikt.2016}. To avoid any influence on the plasma discharge, the admixture was always kept around \SI{10}{ppm}. The low admixture is also important as otherwise an overpopulation of higher rotational states is probable \cite{Bruggeman.2010}. The analysis of the \ce{OH} spectra was realized using the software massiveOES by Vor{\'a}{\v{c}} \textsl{et al.} \cite{Vorac.2017,Vorac.2017b,Vorac.2017c} to fit a synthetic rotational spectrum to the measured one.
	
\begin{figure}[!ht]
	\centering
	\includegraphics[width=0.99\linewidth]{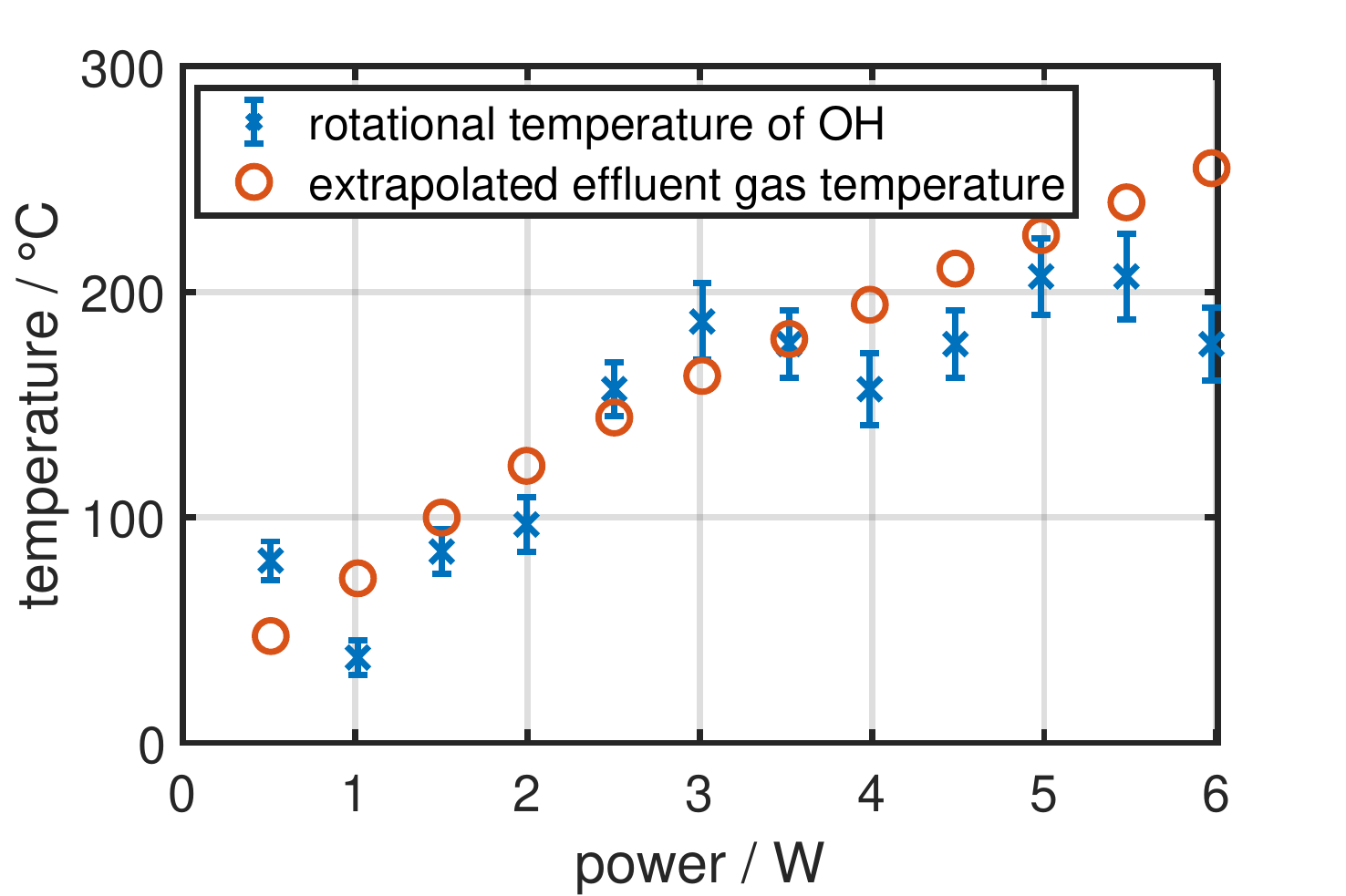}
	\includegraphics[width=0.99\linewidth]{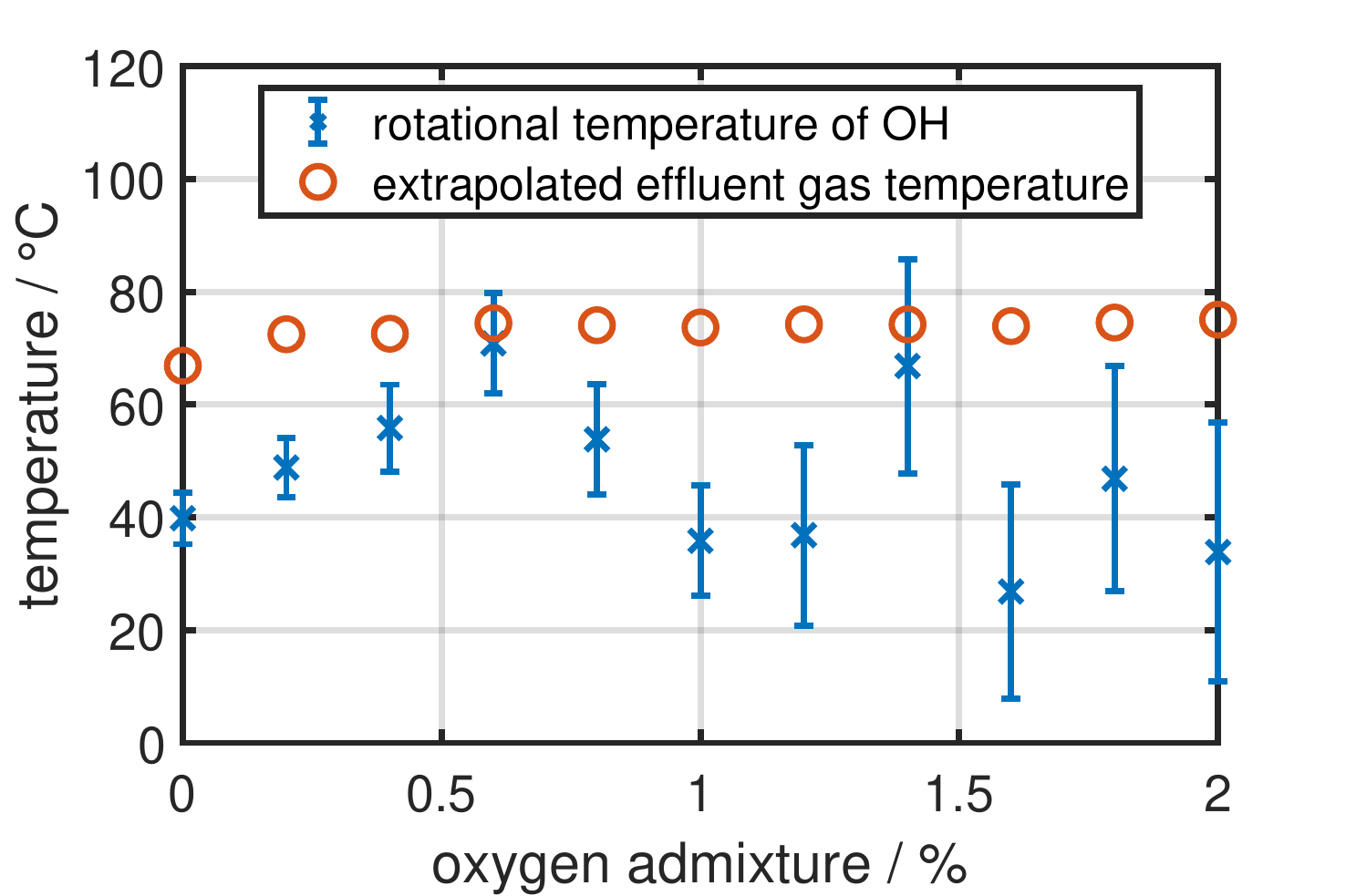}
	\caption{Gas temperature in the plasma extrapolated from the effluent ($\textcolor{orange}{\boldsymbol{\circ}}$) compared to the rotational temperature from the A-X transition of the OH-radical ($\textcolor{blue}{\boldsymbol{\times}}$) as a function the dissipated power at constant admixture of \SI{0.5}{\percent} \ce{O2} (upper figure) and oxygen admixture at constant input power of \SI{1}{W} (lower figure). The rest of the feed gas consisted of \SI{1000}{\sccm} helium and \SI{0.4}{\sccm} argon.}
	\label{fig:heat}
\end{figure}
	
The temperatures obtained using both methods are shown in figure~\ref{fig:heat} for different input powers at constant admixture and different admixtures at constant power. Overall, the gas temperature increases nearly linearly with increasing input power in the observed range from \SI{50}{\degreeCelsius} at \SI{0.5}{\watt} to \SI{250}{\degreeCelsius} at \SI{6}{\watt}. The slope of the linear increase is lower for powers above \SI{3.5}{\watt}. This is probably due to a more efficient heat transfer from the gas to the surroundings at higher temperatures. Also the uncertainty of the rotational temperature increases as the emission intensity of the \ce{OH} system decreases. Overall, the uncertainty of the rotational temperature seems to be underestimated by the fit. The temperatures at low input powers are slightly higher than in the COST-Jet, which is most likely due to the increased length of the plasma channel and therefore the longer residence time of the gas in the plasma \cite{Kelly.2015, Riedel.2020}.

When varying the oxygen admixture at a constant input power of \SI{1}{\watt}, the gas temperature in the plasma stays nearly constant around \SI{75}{\degreeCelsius}. The temperature increase between no admixture and \SI{0.2}{\percent} is probably due to exothermic reactions caused by oxygen radicals and ozone in the effluent \cite{Jeong.2000,Ellerweg.2012}. The increasing uncertainty of the rotational temperature is due to decreasing emission intensity of the OH(A-X) transition. The mean rotational temperature of \ce{OH} is also about \SI{20}{\degreeCelsius} lower than the extrapolated effluent gas temperature. A possible explanation is the position of the measurement. The extrapolation from the effluent gives the gas temperature at the end of the electrodes where the gas has the highest temperature \cite{Kelly.2015}. In contrast, the spectra were recorded at a position about \SI{1}{\centi\meter} before the end of the electrodes, leading to a shorter residence time of the gas in the plasma and therefore lower temperature. The gas temperature under variation of the oxygen admixture for \SI{5}{\watt} input power isn't shown here, because the observed trend is the same as for low input power apart from a higher overall temperature and an even lower \ce{OH} emission intensity leading to a larger uncertainty of the rotational temperature.

The gas temperature measured with the thermocouple is used in the energy balance calculations, as it has a lower uncertainty than the rotational temperature and is a more accurate estimate for the total temperature increase along the plasma channel.

\subsection{Chemistry}
\label{sec:chemistry}

We measured the atomic oxygen density in the discharge using actinometry and molecular beam mass spectrometry. For the analyzed oxygen admixtures, we assume oxygen dissociation being the dominant chemical reaction in the plasma initiated by electrons. Therefore, measuring the density of atomic oxygen gives a good estimation for the power invested in plasma chemical processes. As the subsequent formation of ozone from atomic and molecular oxygen is efficient in oxygen-containing discharges, the ozone density has been studied as well and was included in the estimations for power invested in generation of final chemical products.

\subsubsection{Actinometry}
Actinometry is based on the relation between the emission intensity of an optical transition and the ground-state density of the emitting species. As the emission also depends on the electron energy distribution in the energy range for excitation into the emitting state, a actinometer gas is added to the gas mixture. Argon is used in our case with admixtures as low as \SI{0.04}{\percent}. More details on the theoretical basics and validity range of the method can be found in the literature \cite{Walkup.1986,Katsch.2000, Niemi.2009}. The density of atomic oxygen is calculated via

\begin{equation}
    n_\text{O} = \frac{I_\text{O}}{I_\text{Ar}} \frac{\nu_\text{Ar}}{\nu_\text{O}} \frac{k_\text{e,Ar}^*}{k_\text{e,O}^*} \frac{a_\text{ik,Ar}}{a_\text{ik,O}} n_\text{Ar} - \frac{k_\text{de,\ce{O2}}^*}{k_\text{e,O}^*} n_\text{\ce{O2}} \text{ .}
    \label{eq:oxygen_density}
\end{equation}

The quantities in equation~\ref{eq:oxygen_density} are the emission intensities $I_\text{O}$ of the oxygen transition at \SI{844.6}{\nano\meter} and $I_\text{Ar}$ of the argon transition at \SI{750.4}{\nano\meter}, the frequencies $\nu_\text{O}$ and $\nu_\text{Ar}$ of the transitions, the effective excitation rate coefficients $k_\text{e}^*$, the effective optical branching ratios $a_\text{ik}$, the effective dissociative excitation rate coefficient $k_\text{de}^*$ and the densities $n_\text{Ar}$ and $n_\text{\ce{O2}}$ of argon and molecular oxygen respectively. Calculation of the effective rate coefficients $k^*$ requires knowledge about the averaged electron energy distribution function (EEDF). The determination of the EEDF is beyond the scope of this work, so we used the rate coefficient ratios calculated by \citeauthor{Niemi.2009} \cite{Niemi.2009}. The emission intensities were derived from the spectra measured parallel to the gas flow. The atomic oxygen line lies outside of the wavelength range of the Echelle spectrometer, so it could not be used for this measurement.

\begin{figure}[!ht]
	\centering
	\includegraphics[width=0.99\linewidth]{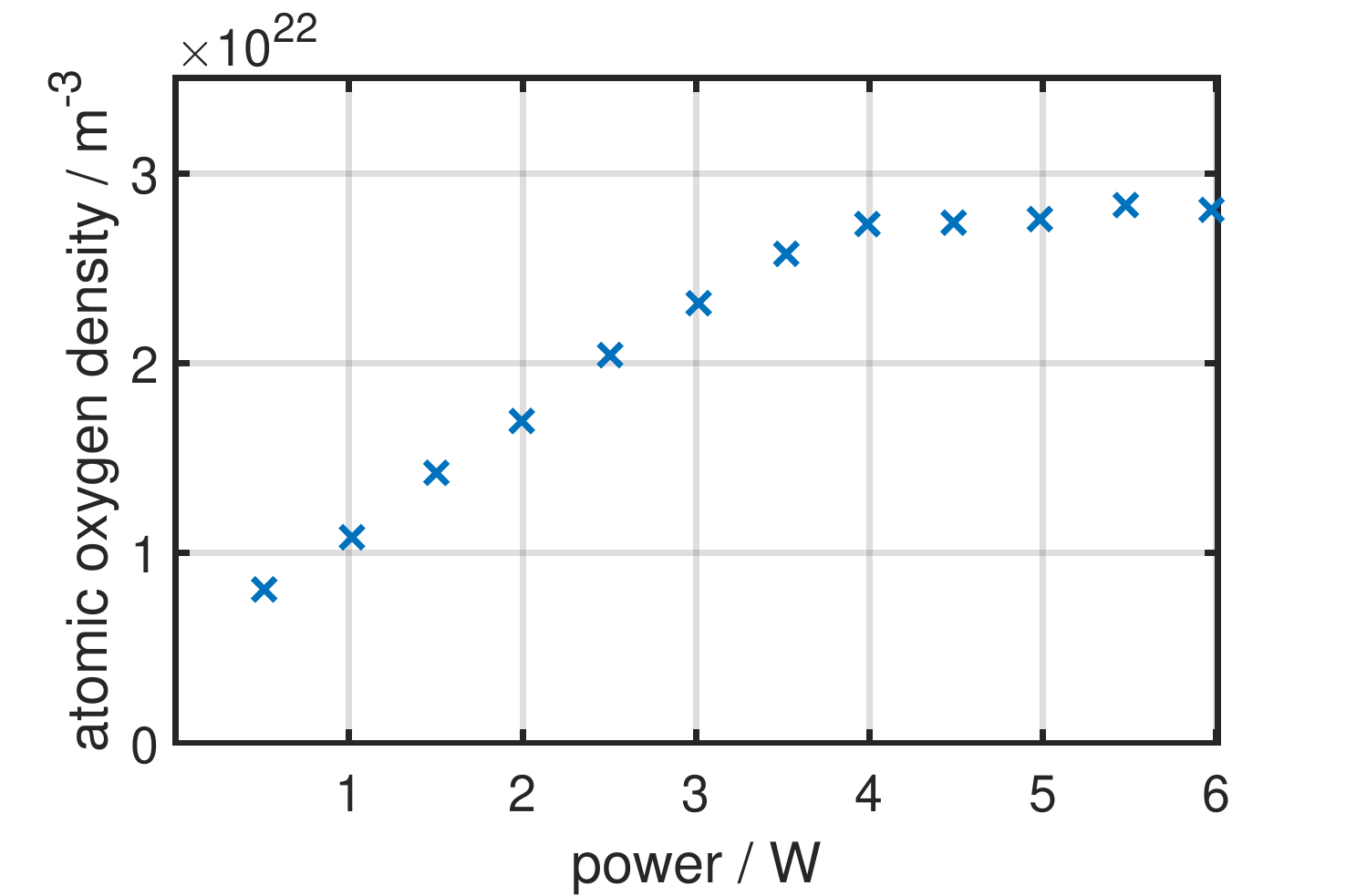}
	\caption{Atomic oxygen density from actinometry as a function of plasma input power. The gas flow consisted of \SI{1000}{\sccm} helium, \SI{0.4}{\sccm} argon and \SI{5}{\sccm} oxygen.}
	\label{fig:density_power_actinometry}
\end{figure}

The upper graph in figure~\ref{fig:density_power} shows the atomic oxygen density calculated from the optical emission spectra as a function of the plasma input power. For powers up to \SI{4}{\watt}, the density rises linearly from \SI{8E21}{\per\cubic\meter} to \SI{2.8E22}{\per\cubic\meter}. Above \SI{4}{\watt}, the slope changes and the density increase is less steep, leading to a density of \SI{2.9E22}{\per\cubic\meter} at \SI{6}{\watt}. Comparable absolute densities have been determined by \citeauthor{Knake.2008} using TALIF and \citeauthor{Niemi.2009} using diagnostic based modeling for similar discharges \cite{Knake.2008,Niemi.2009}. As both authors do not give the plasma input power, direct comparison of the observed trends with our results is difficult. The change in the slope is not induced by the plasma mode change, as this happens at much lower powers (see section~\ref{sec:power}). A possible explanation for the overall sub-linear increase of the density is the linear increase of the electron density with the input power, while the mean electron energy decreases \cite{Lazzaroni.2012}. The oxygen density depends linearly on the electron density but over-linearly on the electron energy \cite{Waskoenig.2010}.

\begin{figure}[!ht]
	\centering
	\includegraphics[width=0.99\linewidth]{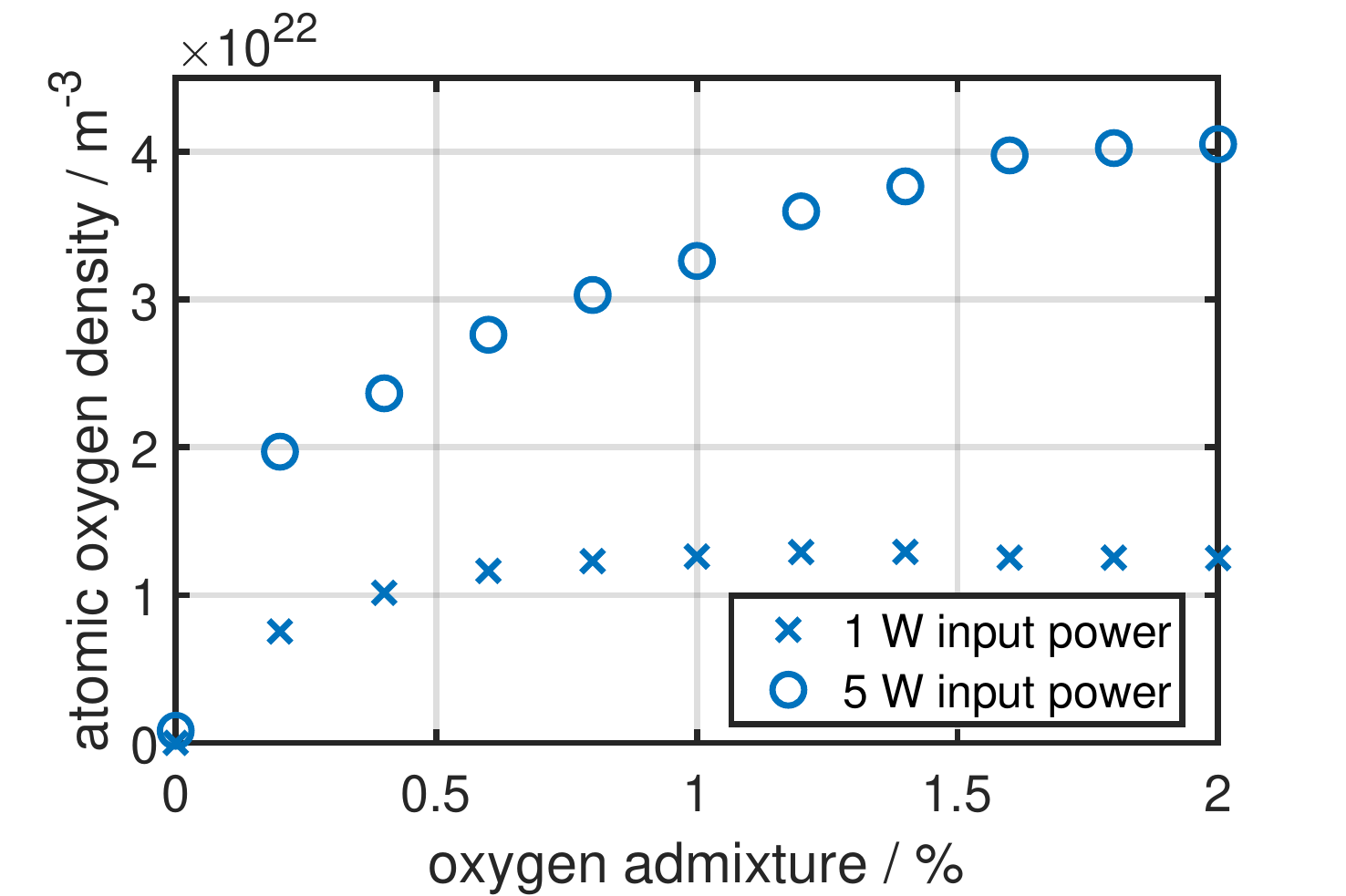}
	\caption{Atomic oxygen density from actinometry as a function of oxygen admixture for constant input power of \SI{1}{W} ($\textcolor{blue}{\boldsymbol{\times}}$, $\alpha$-mode) and \SI{5}{W} ($\textcolor{blue}{\boldsymbol{\circ}}$, $\gamma$-mode). The main gas mixture consisted of \SI{1000}{\sccm} helium and \SI{0.4}{\sccm} argon.}
	\label{fig:density_admixture_actinometry}
\end{figure}

Figure~\ref{fig:density_admixture_actinometry} shows the dependence of the atomic oxygen density measured by actinometry on the oxygen admixture for constant input power of \SI{1}{\watt} ($\textcolor{blue}{\boldsymbol{\times}}$) and \SI{5}{\watt} ($\textcolor{blue}{\boldsymbol{\circ}}$). At \SI{1}{\watt}, the atomic oxygen density shows a steep increase up to \SI{1.1E22}{\per\cubic\meter} at \SI{0.6}{\percent} admixture and then increases much slower to \SI{1.2E22}{\per\cubic\meter} at \SI{1.2}{\percent}. For higher admixtures, the density decreases again to \SI{1.1E22}{\per\cubic\meter} at \SI{2}{\percent}. For an input power of \SI{5}{\watt}, the density rises to \SI{2E22}{\per\cubic\meter} between no admixture and \SI{0.2}{\percent}. Afterwards, a linear increase to \SI{4E22}{\per\cubic\meter} at \SI{1.6}{\percent} admixture is visible. The curve flattens above \SI{1.6}{\percent}. The trend of the atomic oxygen density as a function of the oxygen admixture can be explained by an increasing ozone density. Ozone is produced in three-body-reactions with oxygen atoms and molecules. This has been observed before for atmospheric pressure plasma jets \cite{Ellerweg.2010}.

\subsubsection{MBMS}
We used MBMS as a second method to verify the atomic oxygen densities obtained by actinometry and to measure the ozone density in the effluent.

\begin{figure}[!ht]
	\centering
	\includegraphics[width=0.99\linewidth]{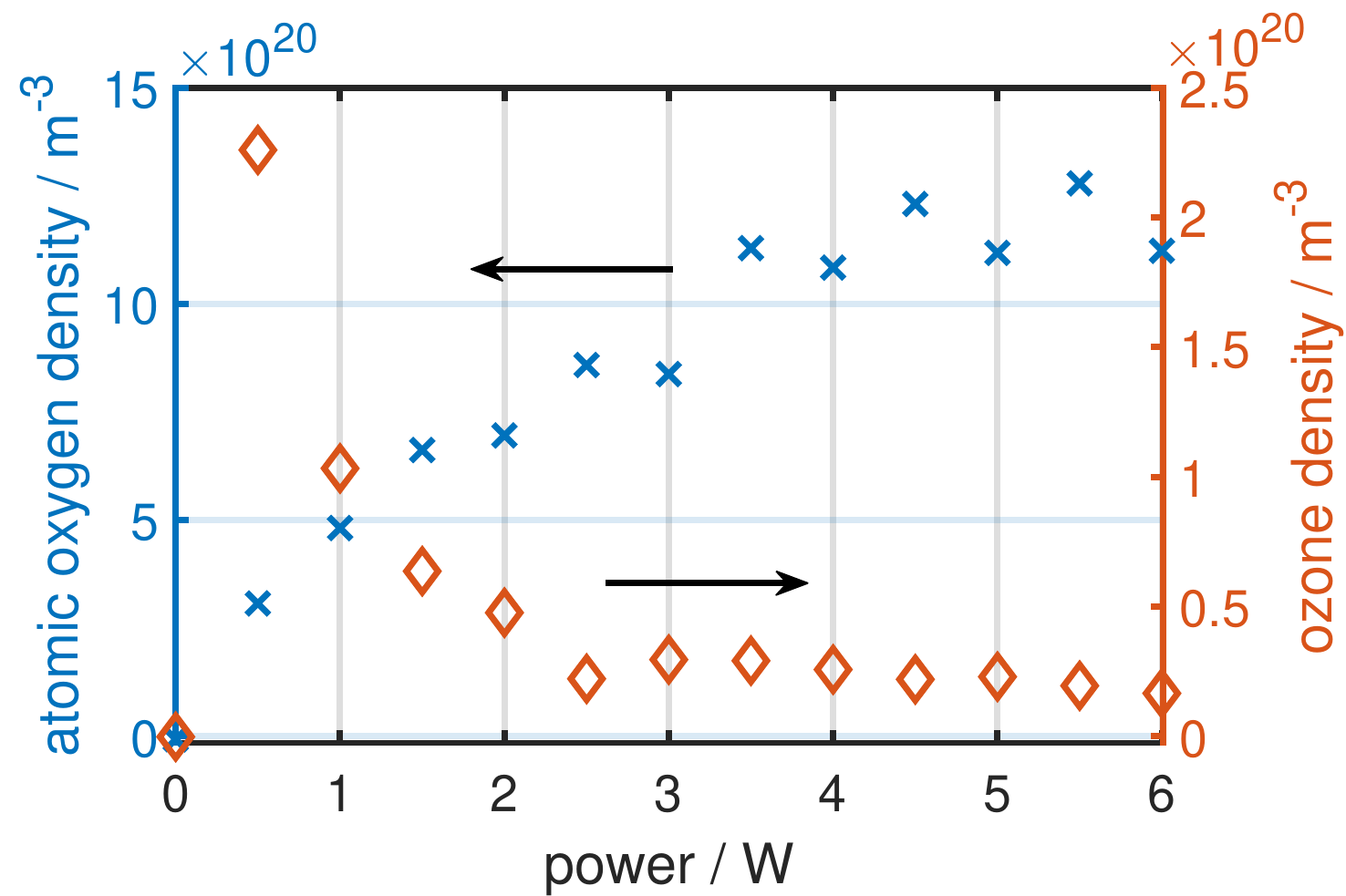}
	\caption{Atomic oxygen density from actinometry as a function of plasma input power. The gas flow consisted of \SI{1000}{\sccm} helium, \SI{0.5}{\sccm} argon and \SI{5}{\sccm} oxygen.}
	\label{fig:density_power}
\end{figure}

The atomic oxygen density under variation of the input power measured by MBMS ($\textcolor{blue}{\boldsymbol{\times}}$) is shown in figure~\ref{fig:density_power}. The atomic oxygen density increases almost linearly with the power up to \SI{7E20}{\per\cubic\meter} at \SI{1.5}{\watt}. Afterwards, the increase gets less steep and the density reaches an approximately constant value of \SI{1.2E21}{\per\cubic\meter} above \SI{5}{\watt}.

The absolute densities are about a factor of \num{20} lower than the values measured by actinometry while showing comparable trends. An explanation is the measurement in the effluent with a distance of \SI{15}{\milli\meter} to the plasma. This leads to a density decrease of a factor of \num{5} to \num{6} compared to inside the plasma as deduced from literature with a controlled helium atmosphere \cite{Ellerweg.2012,Ellerweg.2010} and with an enclosed effluent as it is the case in our setup \cite{Bibinov.2011}. When comparing our results with the given literature, the lower gas flow in our case has to be taken into account. This difference in flow velocity leads to increased atomic oxygen recombination in the effluent. Also, the systematic uncertainty of values measured by MBMS is about a factor of \num{2} resulting from the uncertainty of ionization cross-sections and the density calibration \cite{Ellerweg.2010}. Another explanation is an overestimation of the density by actinometry, most likely due to the use of effective rate coefficients from the literature and the resulting assumptions about the EEDF. Also, actinometry as an optical method always probes the plasma volume where most excitation processes take place. This may also be the volume in which the most chemical reactions happen, so the the obtained atomic oxygen density is not volume-averaged in contrast to the MBMS measurements. In conclusion, the actual atomic oxygen density inside the plasma is expected to be between the density obtained by actinometry as an upper limit and the extrapolated density from MBMS in the effluent as the lower limit.

The ozone density ($\textcolor{orange}{\boldsymbol{\diamond}}$) rises steeply with discharge ignition to \SI{1.8E20}{\per\cubic\meter} at \SI{0.5}{\watt} followed by a fast decrease with rising power to an approximately constant value of \SI{2E19}{\per\cubic\meter} above \SI{3}{\watt}. This decrease is most likely induced by the rising gas temperature (see figure~\ref{fig:heat}). \citeauthor{Ellerweg.2010} measured ozone densities up to \SI{5E20}{\per\cubic\meter} with similar oxygen admixture under discharge voltage variation for the COST-Jet corresponding to the low end of our observed power range \cite{Ellerweg.2010}. The different distances to the plasma of \SI{3}{\milli\meter} in their case and \SI{15}{\milli\meter} in our case, the influence of the capillary and the different gas flows should be taken into account. \citeauthor{Bibinov.2011} measured an ozone density decrease of at least a factor of \num{2} over this distance in a guided effluent due to wall recombination of atomic oxygen and ozone \cite{Bibinov.2011}. Owing to the lower flow velocity and oxygen admixture, the density decrease could be even steeper in our setup, explaining the difference of the ozone densities compared to the COST-Jet. Furthermore, the capillary jet enabled us to perform the measurement over a larger power range, showing, that no considerable change in the ozone density happens above \SI{2}{\watt}.

\begin{figure}[!ht]
	\centering
	\includegraphics[width=0.99\linewidth]{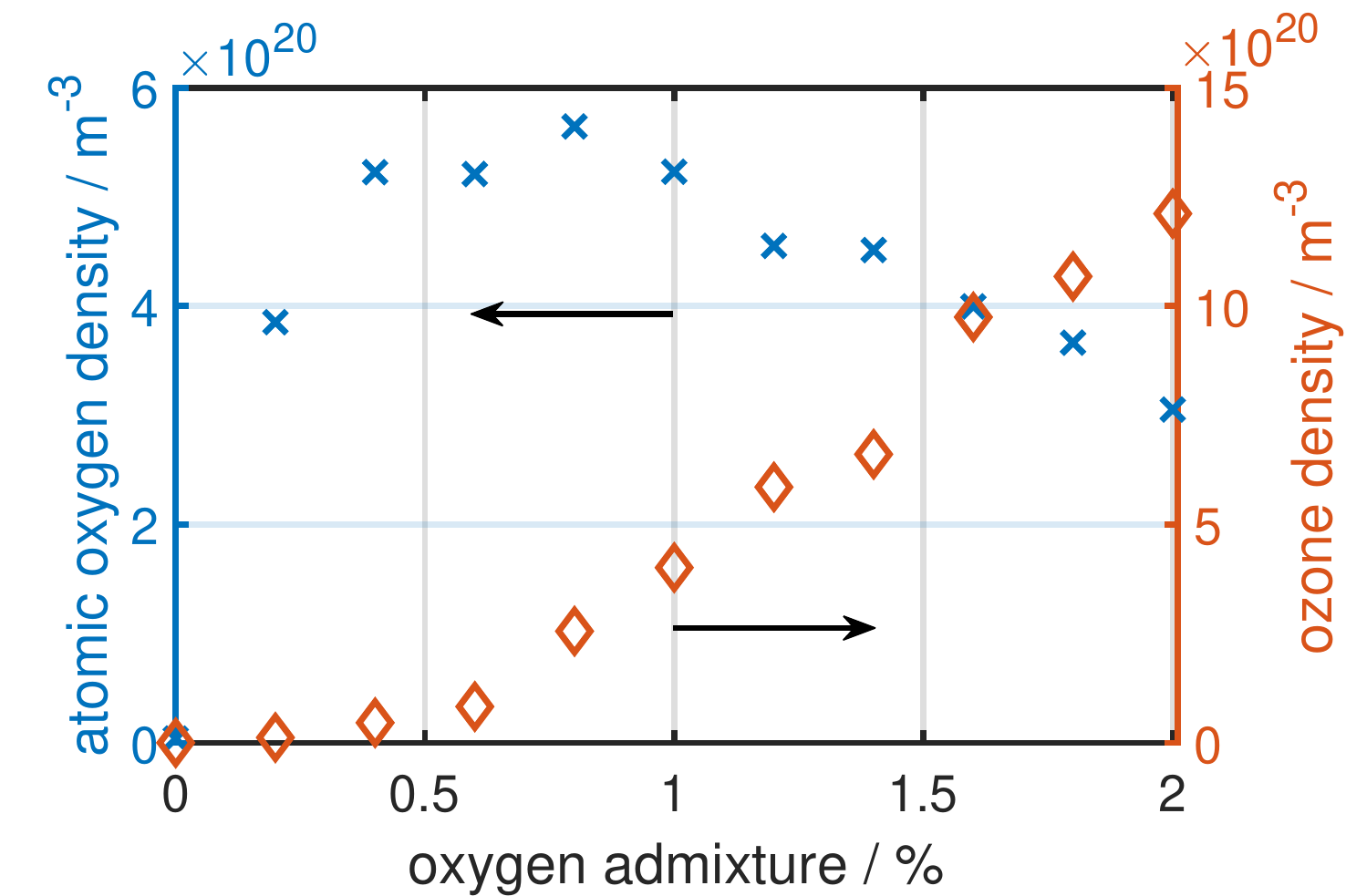}
	\includegraphics[width=0.99\linewidth]{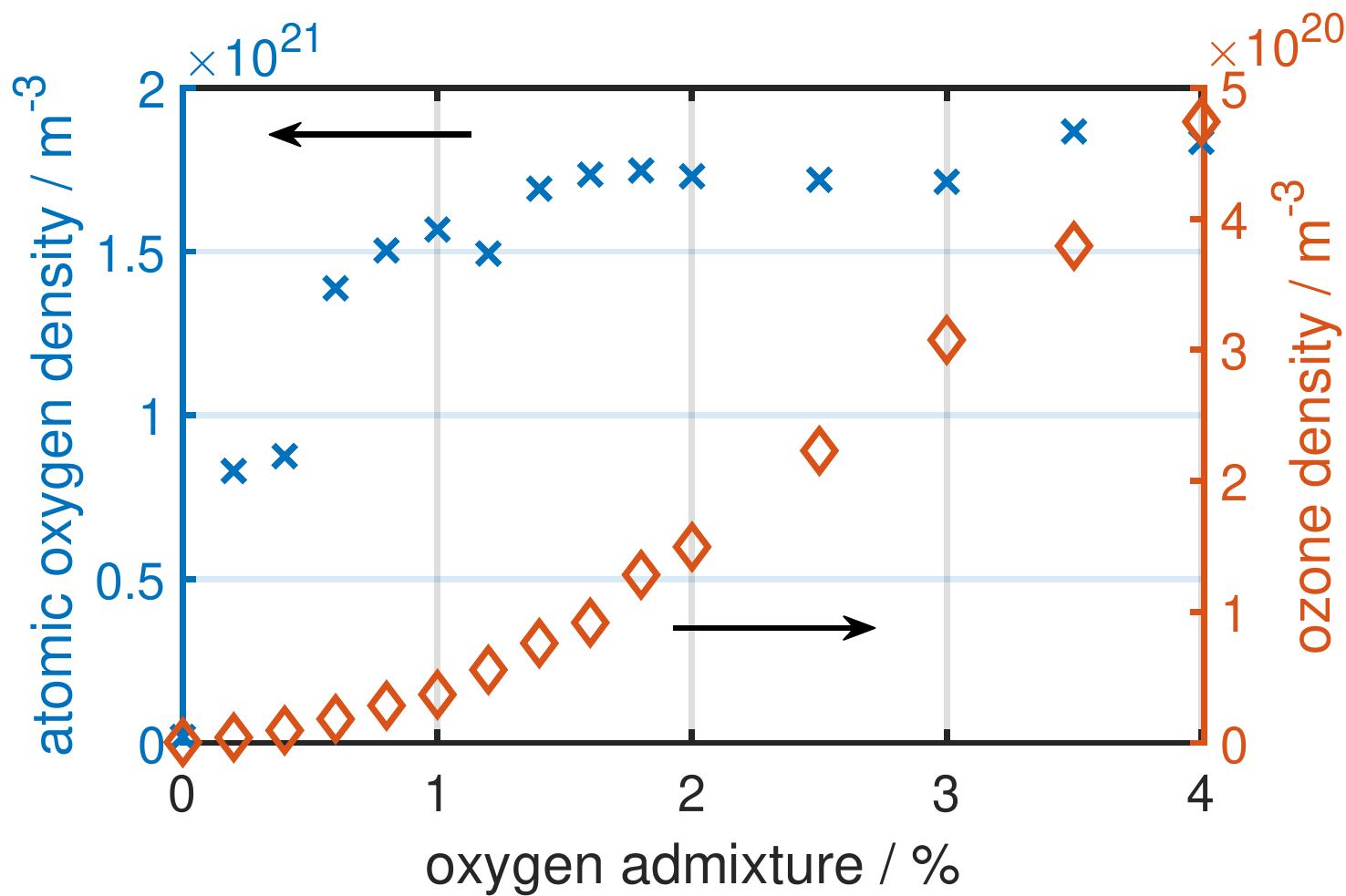}
	\caption{Atomic oxygen ($\textcolor{blue}{\boldsymbol{\times}}$) and ozone ($\textcolor{orange}{\boldsymbol{\diamond}}$) densities measured by MBMS as a function of the oxygen admixture for constant input power of \SI{1}{\watt} and \SI{5}{\watt}. The main gas flow consisted of \SI{1000}{\sccm} helium. Traces of argon were added for MBMS of ozone (\SI{0.5}{\sccm}) as well as neon for MBMS of atomic oxygen (\SI{0.5}{\sccm}).}
	\label{fig:density_admixture_mbms}
\end{figure}

The atomic oxygen density ($\textcolor{blue}{\boldsymbol{\times}}$) as a function of the oxygen admixture measured using MBMS is shown in figure~\ref{fig:density_admixture_mbms}. Again, the input power has been kept constant at \SI{1}{\watt} and \SI{5}{\watt} to study $\alpha$- and $\gamma$-mode behaviour. At \SI{1}{\watt}, the density of atomic oxygen rises steeply for low admixtures up to \SI{5.2E20}{\per\cubic\meter} at \SI{0.4}{\percent}. The curve then flattens and the density shows a maximum of \SI{5.5E20}{\per\cubic\meter} at \SI{0.8}{\percent} before dropping linearly down to \SI{3E20}{\per\cubic\meter} at \SI{2}{\percent}. The absolute densities lie below the values measured at an input power of approximately \SI{0.6}{\watt} in the COST-Jet but the higher distance to the plasma has again to be taken into account \cite{Ellerweg.2010}.

For the input power of \SI{5}{\watt}, the density increases up to \SI{1.75E21}{\per\cubic\meter} at \SI{1.6}{\percent} and then stays constant up to \SI{4}{\percent} admixture. Note that the admixture range exceeds the range for low power and both actinometry measurements. The reason is a modification of the setup (change of mass flow controllers) enabling us to reach higher admixtures. A maximum of the atomic oxygen density in $\gamma$-mode is still not observed, even though \citeauthor{Park.2010} suggested a shift of the maximum observed at low input power to higher admixtures with increasing power based on simulations \cite{Park.2010}. The atomic oxygen density for a plasma jet in $\gamma$-mode and for this admixture range has to our knowledge not been reported before.

The ozone density as a function of the oxygen admixture in figure~\ref{fig:density_admixture_mbms} shows an over-linear increase with no maximum being reached in the observed admixture ranges for both input powers. This increase of the ozone density also explains the decrease of atomic oxygen observed at \SI{1}{\watt} above \SI{0.8}{\percent} admixture. The maximum measured ozone density of \SI{1.2E21}{\per\cubic\meter} is reached at the highest admixture of \SI{2}{\percent}. The ozone density at \SI{5}{\watt} input power and \SI{2}{\percent} admixture is about a factor of \num{10} lower than at low input power. A possible explanation is the overall lower gas density due to the higher gas temperature and increased ozone loss reactions. The highest ozone density of \SI{4.75E20}{\per\cubic\meter} for \SI{5}{\watt} input power is reached at the maximum oxygen admixture of \SI{4}{\percent}. Comparable trends have again been measured by \citeauthor{Ellerweg.2010} in the COST-Jet \cite{Ellerweg.2010}. At \SI{1}{\watt} input power, the absolute ozone densities are about a factor of \num{2} lower than in the COST-Jet. Possible reasons for this difference have been discussed above.

\subsection{Radiation}
\label{sec:radiation}

Inelastic collisions of electrons with other plasma species also lead to excitation of these species. One of the main spontaneous relaxation processes is the emission of radiation. As the measurement of absolute VUV/UV photon fluxes is very challenging and technically demanding, we limited the following measurements to the visible wavelength region using the absolutely calibrated Echelle spectrometer. The emission intensity has been calculated as the sum over all photons per second $I_i$ in a wavelength range $\Delta \lambda_i$ emitted from an effective plasma volume $V_\text{eff}$. This volume takes into account the solid angle from which radiation is detected by the spectrometer. The associated formula is

\begin{equation}
    I_\text{rad} = \frac{V_\text{p}}{V_\text{eff}} \sum_i I_i \Delta \lambda_i \text{ .}
    \label{eq:radiation_intensity}
\end{equation}

Multiplication with the plasma volume $V_\text{p}$ scales the intensity detected from the effective volume up to the overall emitted intensity. In the case of the Echelle spectrometer, the covered wavelength range is \SIrange{200}{780}{\nano\meter}. 

\begin{figure}[!ht]
	\centering
	\includegraphics[width=0.99\linewidth]{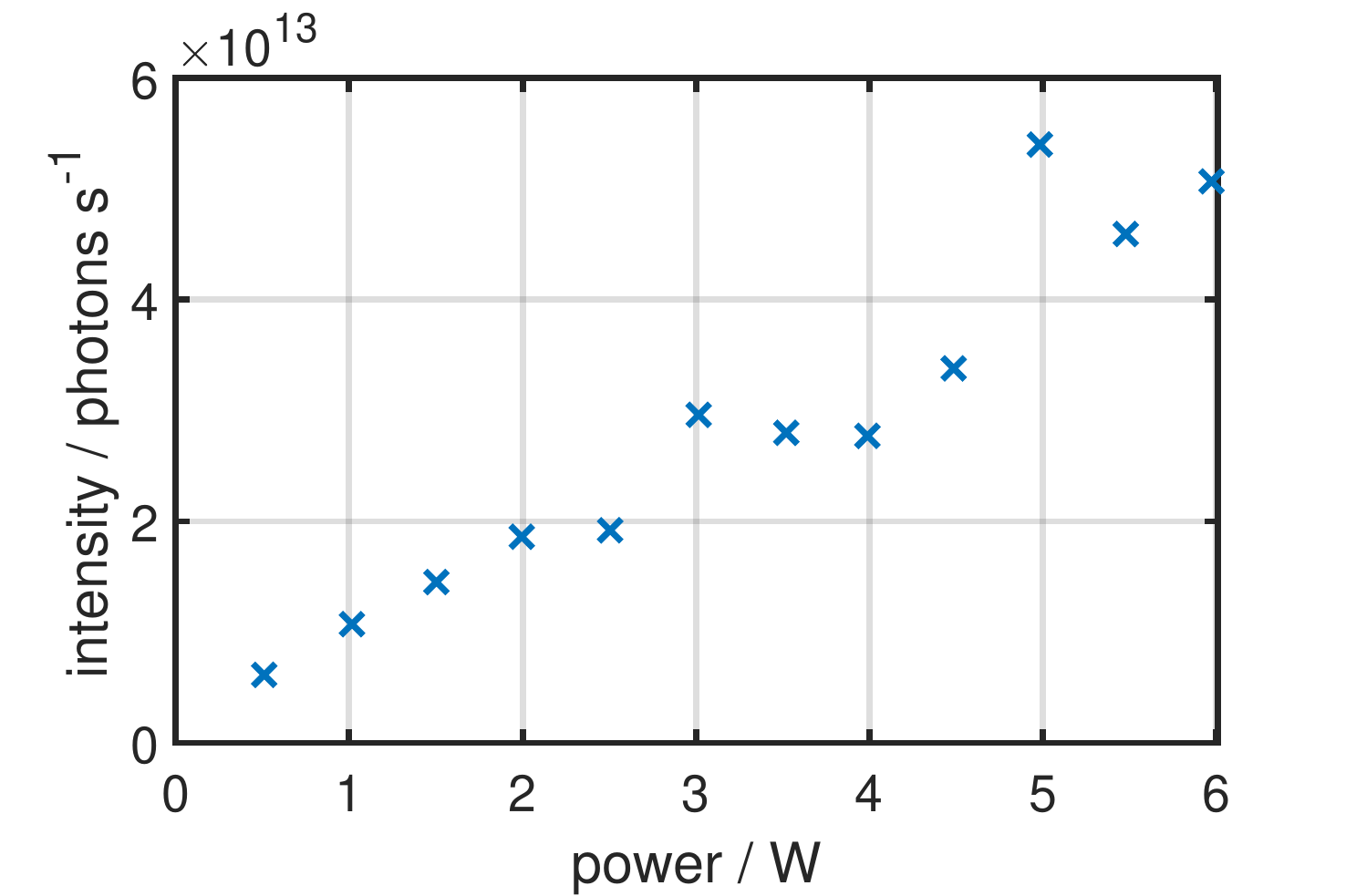}
	\caption{Absolutely calibrated intensity of visible radiation as a function of input power. The gas mixture consisted of \SI{1000}{\sccm} helium, \SI{0.4}{\sccm} argon and \SI{5}{\sccm} oxygen.}
	\label{fig:radpower_power}
\end{figure}

Figure~\ref{fig:radpower_power} shows the intensity in \si{\photons\per\second} emitted in the visible wavelength range as a function of the input power in a range up to \SI{6}{\watt}. The intensity rises approximately linear with the input power from \SI{0.5E13}{\photons\per\second} at \SI{0.5}{\watt} to about \SI{5E13}{\photons\per\second} at \SI{6}{\watt}. This behaviour is expected, since the plasma becomes visibly brighter with rising power.

From the spectrum in figure~\ref{fig:spectrum_actinometry} it is evident, that line radiation dominates. Therefore low emitted power is expected in the measured wavelength range. We also do not expect high radiation power in the VUV/UV-region, as the spectrum is also mostly dominated by line radiation despite of the helium excimer continua. However, the intensity of this continuum radiation drops with oxygen admixture and should be negligible for all admixtures observed in this work \cite{Golda.2020}. If the atomic oxygen lines in the VUV/UV-region made up approximately \SI{1}{\percent} of the input power, the number of excited oxygen atoms would have to be in the order of \SI{1E14}{\per\cubic\meter}. We estimated the density of the excited $^3\text{S}^\circ$-state of oxygen to be around \SI{1E11}{\per\cubic\meter} using a Maxwellian electron energy distribution with \SI{2}{\electronvolt} and cross-sections from \citeauthor{Laher.1990} \cite{Laher.1990}. As the calculated densities of the other excited states from which emission is observed in this wavelength region are orders of magnitude lower than of the $^3\text{S}^\circ$-state under the same conditions, we assume energy loss via VUV/UV-radiation to be negligible. \citeauthor{Jonkers.2002} even omitted energy losses due to radiation in the energy balance for their analytical model because of its small contribution \cite{Jonkers.2002}. Possible sources of uncertainty for the measured powers are the absolute calibration of the spectrometer, the calculation of the effectively observed plasma volume and the gaps in the Echelle-spectrum \cite{Bibinov.2007}. Because of the expected very small contribution of radiation to the overall power balance, we did not calculate the uncertainty here.

\begin{figure}[!ht]
	\centering
	\includegraphics[width=0.99\linewidth]{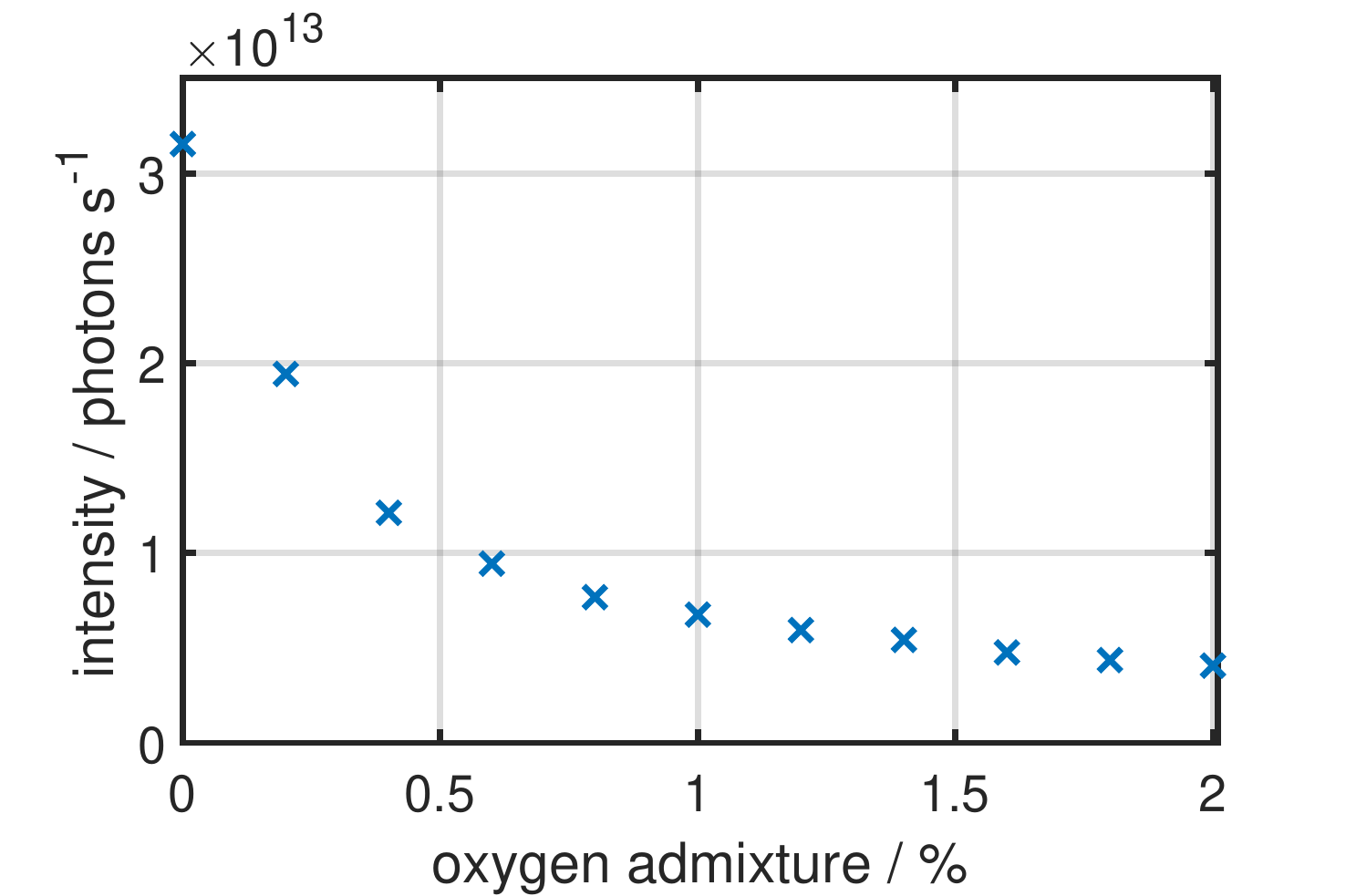}
	\caption{Absolutely calibrated intensity of visible radiation as a function of oxygen admixture for constant input power of \SI{1}{W}. The main gas mixture consisted of \SI{1000}{\sccm} helium and \SI{0.4}{\sccm} argon.}
	\label{fig:radpower_admixture}
\end{figure}

The emission intensity as a function of the oxygen admixture up to \SI{2}{\percent} is shown in figure~\ref{fig:radpower_admixture} for a constant input power of \SI{1}{\watt}. The highest intensity of \SI{3.1E13}{\photons\per\second} is measured for no oxygen admixture. The intensity then drops steeply to just below \SI{1E13}{\photons\per\second} at \SI{0.6}{\percent} admixture and then decreases slowly to \SI{0.4E13}{\photons\per\second} at \SI{2}{\percent}. Even though the plasma input power has been kept constant for all admixtures by increasing the applied voltage, the plasma became darker with rising admixture. This qualitatively confirms the observed trend of the intensity. As stated above, we expect the radiation in the VUV/UV-region to follow the same trend with negligibly small absolute values. We also measured the radiation power for various admixtures at a constant power of \SI{5}{\watt} and observed the same trend with higher absolute values as expected from the input power variation.

\subsection{Discussion of the energy balance}
\label{sec:discussion}

\begin{figure}[!ht]
	\centering
	\includegraphics[width=0.99\linewidth]{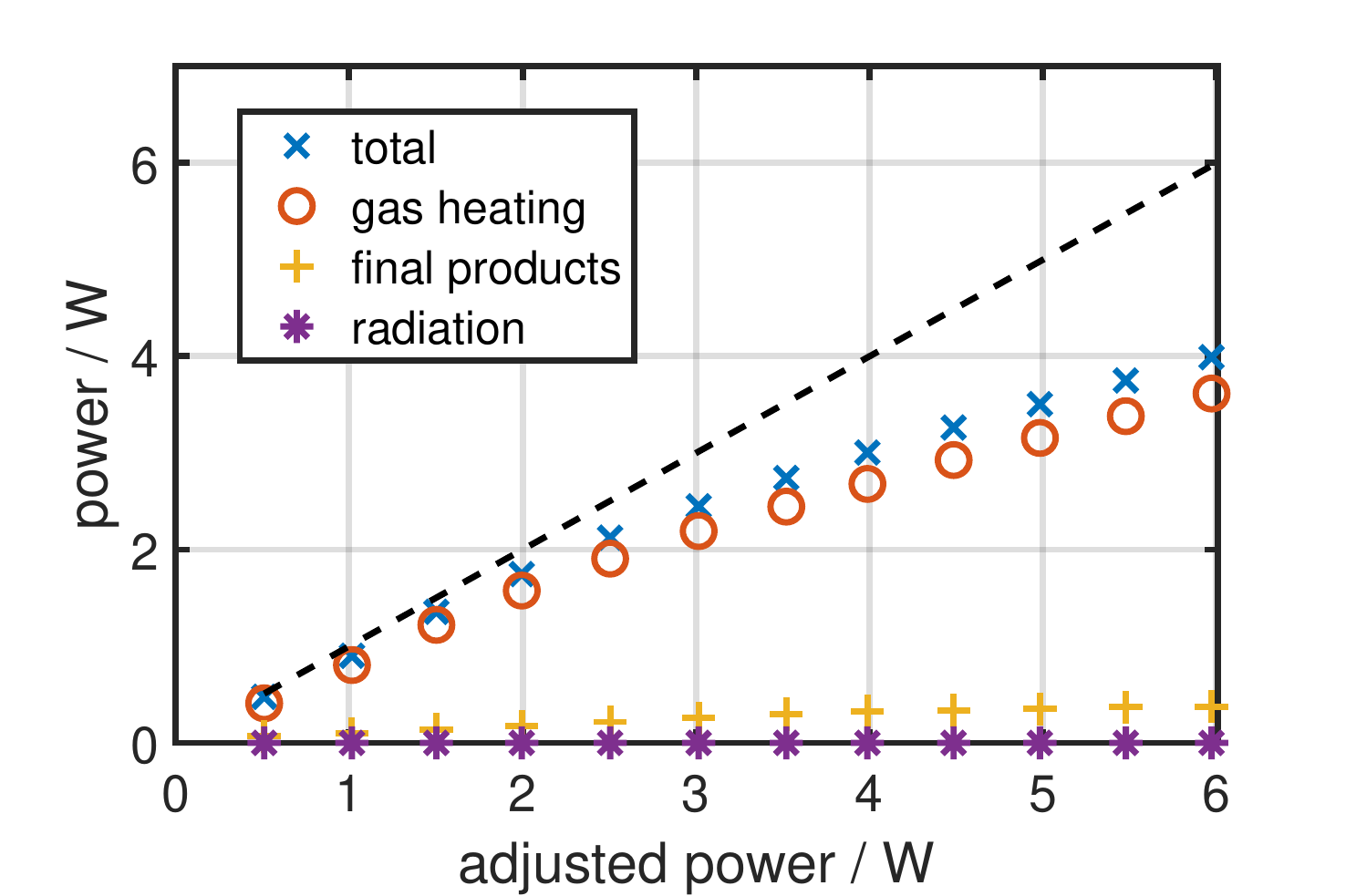}
	\caption{Power balance of the capillary plasma jet as a function of the input power. The gas mixture consisted of \SI{1000}{sccm} helium, \SI{0.4}{sccm} argon and \SI{5}{sccm} oxygen.}
	\label{fig:powerbalance_power}
\end{figure}

\begin{figure}[!ht]
	\centering
	\includegraphics[width=0.99\linewidth]{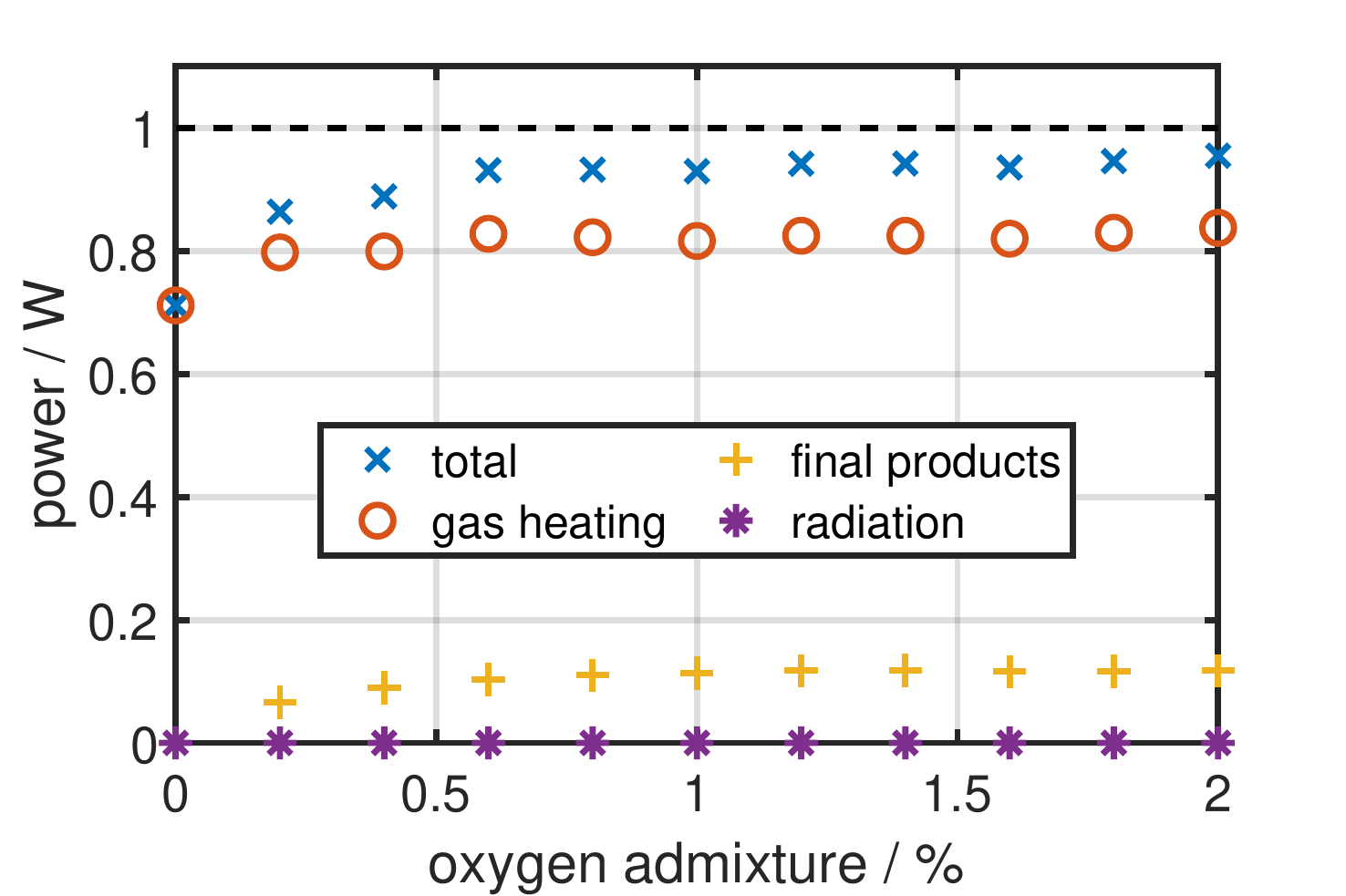}
	\caption{Power balance of the capillary plasma jet as a function of the oxygen admixture for constant input power of \SI{1}{W}. The rest gas flow consisted of \SI{1000}{sccm} helium and \SI{0.4}{sccm} argon.}
	\label{fig:powerbalance_admixture}
\end{figure}

Using the the plasma input power as a reference, we were able to combine the results of the presented measurements to obtain a power balance of the plasma in the capillary jet under variation of the external parameters power (figure~\ref{fig:powerbalance_power}) and oxygen admixture (figure~\ref{fig:powerbalance_admixture}). The three measured contributions to the overall dissipated power (dashed line) are heat ($\textcolor{orange}{\boldsymbol{\circ}}$), generation of atomic oxygen and ozone as final chemical products ($\textcolor{yellow}{\boldsymbol{+}}$) and radiation in the visible wavelength range ($\textcolor{violet}{\boldsymbol{\ast}}$). The sum of the contributions is also shown ($\textcolor{blue}{\boldsymbol{\times}}$). If the power balance was complete, the sum of the contributions would give the input power (dashed line).

To calculate the power put into heating of the feed gas, we took into account the heat

\begin{equation}
    Q = c_\text{He} m \Delta T
    \label{eq:heat}
\end{equation}

stored in the helium fraction of the gas and the residence time

\begin{equation}
    t = \frac{p}{p_\text{s}} \frac{T_\text{s}}{T} \frac{V_\text{p}}{\Phi_\text{s}}
    \label{eq:residence_time}
\end{equation}

of the gas in the plasma channel that can be calculated from the gas flow $\Phi_\text{s}$ under standard conditions. The other quantities are the heat capacity of helium $c_\text{he}$, the mass of helium in the plasma volume at the current gas temperature $m$, the temperature change $\Delta T$, the pressure $p$, the current temperature $T$ (under standard conditions $p_\text{s}$ and $T_\text{s}$) and the plasma volume $V_\text{p}$. The resulting equation for the power is

\begin{equation}
    P_\text{heat} = \frac{Q}{t} = c_\text{he} m_\text{he} \Delta T \frac{p_\text{s} \Phi_\text{s}}{k_\text{B} T_\text{s}}
    \label{eq:power_heat}
\end{equation}

with the atomic weight of helium $m_\text{he} = \SI{4}{u}$ and the Boltzmann constant $k_\text{B}$.

From the measured atomic oxygen and ozone densities, we calculated the fraction of the power which is dissipated by the electrons in inelastic collisions with oxygen molecules leading to oxygen dissociation and subsequently to ozone production. The atomic oxygen densities obtained by actinometry were used here to estimate the maximum power put into oxygen dissociation. This may be an overestimation for this specific reaction, but as we neglected other chemical reactions, e.g. with impurities like nitrogen or water, it may be a good estimation for the overall power put into plasma chemistry. The energy $E_\text{chem}$ needed to create the number of oxygen atoms and ozone molecules in the discharge volume $V_\text{p}$ is calculated via

\begin{equation}
    E_\text{chem} = \left( n_\text{O} \frac{E_\text{d}}{2} + n_\text{\ce{O3}} E_\text{f} \right) \frac{V_\text{p}}{N_\text{A}}  \text{ .}
    \label{eq:dissociation_energy}
\end{equation}

The other quantities in equation~\ref{eq:dissociation_energy} are the density of atomic oxygen $n_\text{O}$, the Avogadro-constant $N_\text{A}$, the dissociation energy $E_\text{d}$ of molecular oxygen and the formation energy $E_\text{f}$ of ozone in \si{\joule\per\mol}. Dividing this value by the residence time $t$ of the species in the plasma (equation~\ref{eq:residence_time}) gives the dissipated power

\begin{equation}
    P_\text{chem} = \left( n_\text{O} \frac{E_\text{d}}{2} + n_\text{\ce{O3}} E_\text{f} \right) \frac{1}{N_\text{A}} \frac{p_\text{s}}{p} \frac{T}{T_\text{s}} {\Phi_\text{s}} \text{ .}
\end{equation}

Using the number of photons per second $I_i$ in a wavelength range $\Delta \lambda_i$, the power dissipated by radiation

\begin{equation}
    P_\text{rad} = \frac{V_\text{p}}{V_\text{eff}} hc \sum_i I_i \frac{\Delta \lambda_i}{\lambda_i}
    \label{eq:radiation_power}
\end{equation}

can be calculated.

With variation of the plasma input power in a range from \SI{0.5}{\watt} to \SI{6}{\watt}, we were able to study the influence of the discharge modes on the energy transport inside the plasma. The power balance shows that the largest part of the input power is converted to heat. This holds true for all measured input powers. Like the gas temperature, the power dissipated by gas heating increases linearly with the input power. At the same time, the relative percentage decreases from nearly \SI{100}{\percent} between \SI{0.5}{\watt} and \SI{1.5}{\watt} to approximately \SI{60}{\percent} at \SI{6}{\watt}.

A possible explanation for this decrease is the more efficient heat transfer via conduction, radiation and convection to the surroundings with rising temperature difference. We investigated this assumption by measuring the temperature of the stainless steel shielding around the electrodes at different positions with an IR-camera and a thermocouple. Using the average of the obtained temperatures in a steady state, we estimate the power dissipated by heat radiation of the setup to be roughly \SI{10}{\percent} of the input power. This fraction was the same for \SI{1}{\watt} and \SI{5}{\watt} input power. The remaining missing part in the energy balance is very probably lost to heating of the surrounding air through conduction and convection.

The power dissipated creating the chemical products atomic oxygen and ozone also increases with higher input power to a maximum value just below \SI{0.5}{\watt}. Therefore we conclude, that the initiation of plasma chemical processes consumes at most \SI{10}{\percent} of the total input power. This is a large fraction considering that only \SI{0.5}{\percent} \ce{O2} have been admixed to the feed gas. The most atomic oxygen per \si{\watt} input power is produced at \SI{0.5}{\watt}, which is promising for applications relying on energy efficiency or low gas temperatures.

Only a negligible amount of power is channeled into radiation. The values for the radiation power are of the order of \SI{1E-14}{\watt}. \citeauthor{Moravej.2004} calculated the emitted power density of an atmospheric pressure discharge with an electron temperature of $T_\text{e} = \SI{2}{\electronvolt}$ and an electron density of $n_\text{e} = \SI{1E17}{\per\cubic\meter}$ to be approximately \SI{4E-6}{\watt\per\cubic\meter} \cite{Moravej.2004}. The values for the electron temperature and density are in a realistic range for a discharge similar to the capillary jet \cite{Waskoenig.2010}. Taking into account the plasma volume in this study, we obtain power densities around \SI{2.5E-7}{\watt\per\cubic\meter}. Due to the uncertainties in the calculation of the emission intensity and power, a difference of one order of magnitude is still a good agreement.

The variation of the oxygen admixture shows the same distribution of the input power between the three measured contributions (figure~\ref{fig:powerbalance_admixture}). This distribution seems basically independent of the molecular oxygen admixture in the observed range. This holds true for admixtures higher than \SI{0.5}{\percent}. Below this admixture, the power dissipated via heat and chemistry decreases. This is probably because of chemical processes being dominated by impurities like nitrogen which cannot be quantified with the used diagnostics. Additionally, with no oxygen admixture, VUV/UV-continuum radiation from helium excimers can make a significant contribution to the overall power balance which is not captured by our measurements.

\section{Conclusion}
\label{sec:conclusion}

\begin{figure}[!ht]
	\centering
	\includegraphics[width=0.99\linewidth]{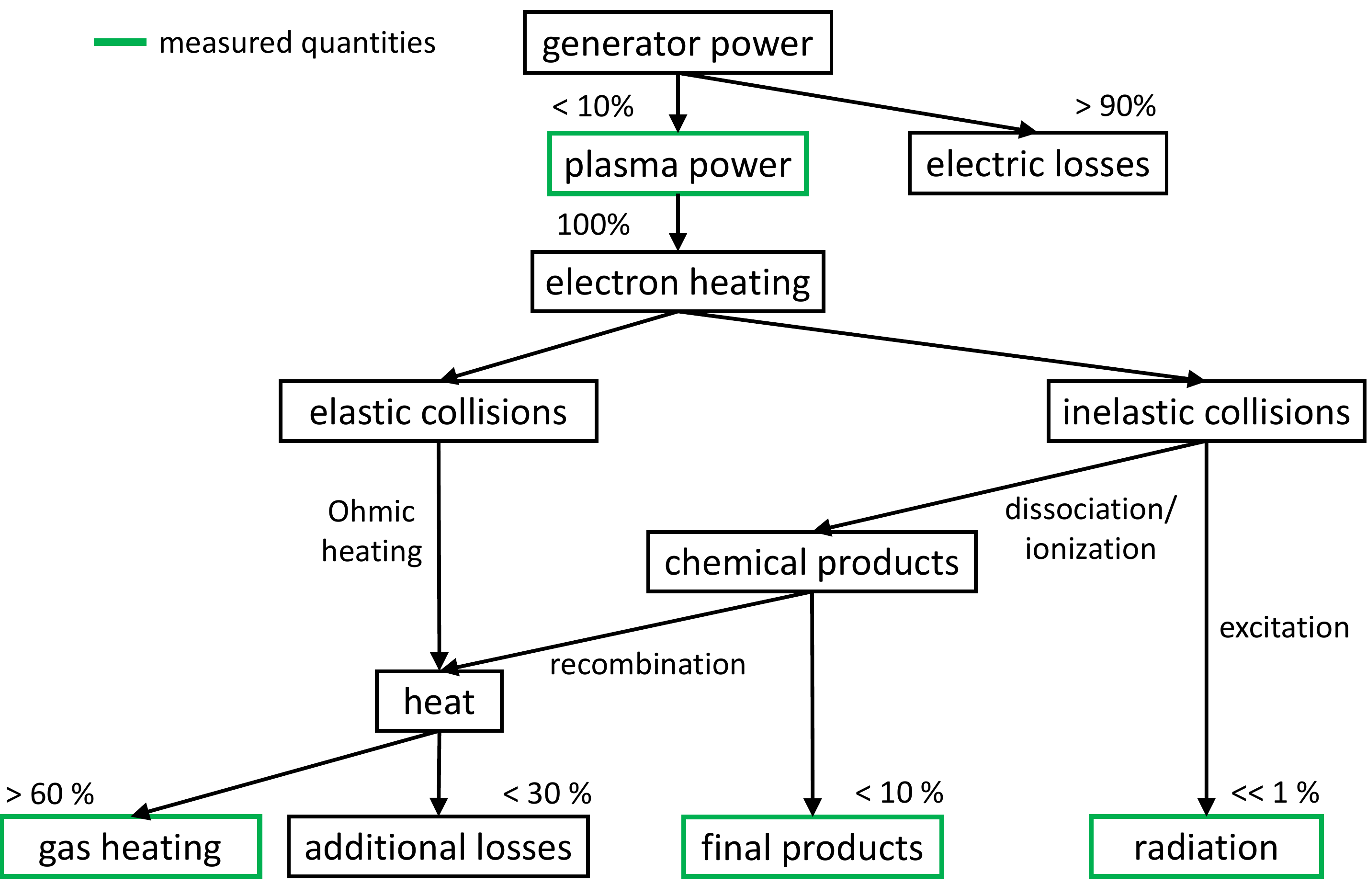}
	\caption{Energy transport diagram of the atmospheric pressure RF-driven capillary plasma jet.}
	\label{fig:powerbalance_diagram}
\end{figure}

The energy transport in the atmospheric pressure rf-driven capillary plasma jet is summarised in the diagram shown in figure~\ref{fig:powerbalance_diagram}. The power measurement enabled us to estimate the fraction of the generator power that is coupled into the plasma, to be below \SI{10}{\percent}. This is confirmed by results from \citeauthor{Stewig.2021} for a comparable plasma device featuring a capillary as dielectric \cite{Stewig.2021}. The power is selectively coupled to the electrons and distributed between the plasma species via elastic and inelastic collisions. Between \SI{60}{\percent} and \SI{100}{\percent} of the power are converted to heat while less than \SI{10}{\percent} is used to initiate chemical processes. Power losses due to radiation make up a negligible amount of the dissipated power. We identified \SIrange{0}{30}{\percent} of the dissipated power to be additional losses which were not quantified. These losses are most likely due to heat transfer from the gas to the setup and surrounding air. For this reason, active cooling of the setup would lead to lower gas temperatures when needed for specific applications. Using a shorter plasma channel or higher gas flow may also reduce the gas temperature but might have an influence on the generation of atomic oxygen as well. Overall, the parameters power and oxygen admixture do not substantially change the energy transport into the different channels. To further explore the possibility of energy transport control, an excitation voltage waveform or frequency variation might be promising via influencing the EEDF.
	
\section{Acknowledgements}

The authors would like to thank Luka Hansen and Holger Kersten from the Plasma Technology Group at Kiel University for help with the temperature measurements. Volker Rohwer, Michael Poser and Mario Knüppel are gratefully acknowledged for technical support. This work was partially funded by the DFG within the project \textit{Cold atmospheric plasmas for material synthesis: production of silicon quantum dots and particle free thin-film deposition} (project-ID BE 4349/7-1) and the project \textit{PlasNOW - Plasma generated Nitric Oxide in Wound healing} (project-ID  SCHU 2353/9-1). JG acknowledges funding by the Faculty of Mathematics and Natural Sciences, Kiel University, Germany, in the framework of the Award for Young Women Scientists 2019.

\printbibliography

\end{document}